\documentclass[
  aps,
  prb,
  reprint,
  superscriptaddress, 
  twocolumn]{revtex4-2}

\usepackage[utf8]{inputenc}
\usepackage[sectionbib]{chapterbib}
\usepackage{amsmath}
\usepackage{amssymb}
\usepackage{bm}
\usepackage[dvipsnames]{xcolor}
\usepackage[
  colorlinks=true,
  linkcolor=Blue,
  citecolor = Blue,
  urlcolor = Blue,
  unicode
  ]{hyperref}
\usepackage{url}
\usepackage{appendix}
\usepackage{graphicx}
\usepackage{physics}
\usepackage{siunitx}
\usepackage[nameinlink]{cleveref}
\crefname{figure}{Fig.}{Figs.}
\Crefname{figure}{Figure}{Figures}
\crefname{equation}{Eq.}{Eqs.}
\Crefname{equation}{Equation}{Equations}

\begin{document}
\title{Wide-field quantitative magnetic imaging of superconducting vortices using perfectly aligned quantum sensors}
\author{Shunsuke Nishimura}
\affiliation{Department of Physics, The University of Tokyo, Bunkyo-ku, Tokyo, 113-0033, Japan}
\author{Taku Kobayashi}
\affiliation{Department of Physics, The University of Tokyo, Bunkyo-ku, Tokyo, 113-0033, Japan}
\author{Daichi Sasaki}
\affiliation{Department of Physics, The University of Tokyo, Bunkyo-ku, Tokyo, 113-0033, Japan}
\author{Takeyuki Tsuji}
\affiliation{Department of Electrical and Electronic Engineering, School of Engineering, Tokyo Institute of Technology, Meguro, Tokyo 152-8552, Japan}
\author{Takayuki Iwasaki}
\affiliation{Department of Electrical and Electronic Engineering, School of Engineering, Tokyo Institute of Technology, Meguro, Tokyo 152-8552, Japan}
\author{Mutsuko Hatano}
\affiliation{Department of Electrical and Electronic Engineering, School of Engineering, Tokyo Institute of Technology, Meguro, Tokyo 152-8552, Japan}
\author{Kento Sasaki}
\affiliation{Department of Physics, The University of Tokyo, Bunkyo-ku, Tokyo, 113-0033, Japan}
\author{Kensuke Kobayashi}
\affiliation{Department of Physics, The University of Tokyo, Bunkyo-ku, Tokyo, 113-0033, Japan}
\affiliation{Institute for Physics of Intelligence, The University of Tokyo, Bunkyo-ku, Tokyo, 113-0033, Japan}
\affiliation{Trans-scale Quantum Science Institute, The University of Tokyo, Bunkyo-ku, Tokyo, 113-0033, Japan}
\date{\today}

\begin{abstract}
Various techniques have been applied to visualize superconducting vortices, providing clues to their electromagnetic response.
Here, we present a wide-field, quantitative imaging of the stray field of the vortices in a superconducting thin film using perfectly aligned diamond quantum sensors.
Our analysis, which mitigates the influence of the sensor inhomogeneities, visualizes the magnetic flux of single vortices in YBa$_2$Cu$_3$O$_{7-\delta}$ with an accuracy of $\pm10$~\%.
The obtained vortex shape is consistent with the theoretical model, and penetration depth and its temperature dependence agree with previous studies, proving our technique's accuracy and broad applicability. 
This wide-field imaging, which in principle works even under extreme conditions, allows the characterization of various superconductors.
\end{abstract}

\maketitle

Superconducting vortex, as a manifestation of macroscopic quantum effects, is one of the central subjects in the physics of superconductivity.
Diverse vortex phases such as vortex lattice, vortex liquid, and Bragg glass appear in type-II superconductors' mixed state~\cite{fisher1991thermal,blatter1994vortices,shibata2002phase}. 
Those phases and vortex dynamics lead to bulk electromagnetic responses of superconductors and thus have been under vigorous investigation. 
Besides, since the flux quantization in superconducting vortices originates from the gap symmetry, anomalous quantization such as a half-quantum vortex in $p$--wave superconductors~\cite{salomaa1987quantized,volovik1999monopole} is proposed to emerge as a signature of unconventional pairing symmetry. 
Therefore, techniques that can quantitatively image quantum vortices under various temperatures, pressures, and magnetic fields would help probe a wide variety of superconductivity with open questions.

Several techniques are available to visualize local magnetic fields~\cite{bending1999,celotta2012,marchiori2022nanoscale,scholten2021widefield}.
In particular, scanning techniques using sensor chips are widely used for quantitative measurements of magnetic flux density~\cite{kirtley1999scanning,finkler2012scanning,kirtley2010fundamental,marchiori2022nanoscale}.
In such scanning techniques, superconducting quantum interference devices (SQUIDs)~\cite{kirtley1999scanning,finkler2012scanning} and nitrogen-vacancy (NV) centers in diamonds~\cite{thiel2016quantitative,pelliccione2016scanned} are prominent as sensors.
While SQUIDs have excellent sensitivity, NV centers operate under severe environments such as high temperatures and high magnetic fields~\cite{schirhagl2014nitrogen,fu2020sensitive}.
Scanning microscopy provides nanoscale spatial resolution and high accuracy~\cite{kirtley2010fundamental}.
As for the NV-center technique, alternatively, imaging with a wide field of view exceeding $(100~\si{\micro\meter}\times 100~\si{\micro\meter})$ is possible with a camera and NV ensemble sensors~\cite{levine2019principles,scholten2021widefield}.
This technique is beneficial in terms of high throughput~\cite{scholten2021widefield}. Furthermore, it can be introduced into extreme environments such as ultrahigh pressure~\cite{Hsieh2019,Lesik2019,Toraille2020}, which are not accessible by the scanning technique. Thus, it aids in researching novel superconductors at high temperatures and pressures~\cite{drozdov2015conventional,mozaffari2019superconducting}. 
Using this technique, efforts have been made particularly to image the stray magnetic fields of superconducting quantum vortices~\cite{schlussel2018wide, scott2020}, but achieving magnetic accuracy close to the scanning technique~\cite{thiel2016quantitative} has been challenging. The issue arises primarily due to the fact that the measurement of superconductors is conducted in a low magnetic field, where the inhomogeneity of the sensor's strain parameter~\cite{rondin2014magnetometry} and the signal overlap resulting from a diamond sensor ensemble with four NV axes render quantitative analysis to extract field component perpendicular to the superconductors' surface practically impossible.

Here, we address these issues by utilizing a perfectly aligned NV ensemble sensor~\cite{miyazaki2014atomistic,tahara2015quantifying,ishiwata2017perfectly,ozawa2017formation,Tsuji2022} and implementing an analysis that eliminates sensor inhomogeneities resulting from strain distribution, complemented by reference measurements in a zero magnetic field. Consequently, we report a quantitative wide-field magnetic imaging of superconducting vortices in a thin film of a typical high-$T_c$ superconductor YBa$_2$Cu$_3$O$_{7-\delta}$ (YBCO). The combination of the inherent high throughput of widefield NV microscopy and achieved quantitativeness enables statistical analysis. The obtained statistics is consistent with the single quantization of vortices. Moreover, the stray magnetic field distribution aligns well with theoretical models, offering an alternative method for estimating the magnetic field penetration depth.
Our technique, which combines high throughput and accuracy, is helpful for comprehensive characterization, including exploring unconventional superconductors~\cite{stewart2017unconventional,sarma2006proposal,how2020half}.

\begin{figure}
\begin{center}
\includegraphics{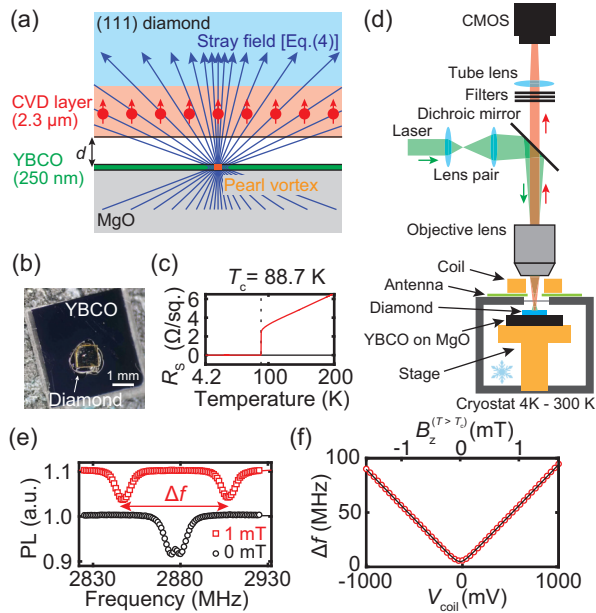}
\caption{
Overview of the NV ensemble-based magnetic imaging.
(a) Schematic of the vortex stray field imaging.
(b) Photograph of the sample.
The diamond chip and the YBCO thin film are bonded by varnish.
(c) Temperature dependence of the sheet resistance of the YBCO thin film.
The horizontal solid black line indicates zero resistance.
(d) Schematic of the microscope system.
A magnetic shield surrounds the cryostat
and the objective lens (not shown).
(e) Typical ODMR spectra.
The black circles and red squares are data acquired at external fields of 0~mT and 1~mT, respectively.
An offset of 0.1 is added to the 1~mT data for visibility.
(f) The coil voltage $V_\text{coil}$ dependence of the splitting between resonance frequencies $\Delta f$.
The upper axis is the magnetic field $B_z^{(T>T_c)}$ estimated from the fitting (solid black line).
This data is acquired at $90~\si{K}~(>T_c)$.
\label{fig1}
}\end{center}
\end{figure}

We use NV ensemble sensors at the diamond surface to visualize vortex stray magnetic fields.
\Cref{fig1}(a) is the measurement schematic.
The sensors are located in a thin film grown on a (111) Ib diamond substrate ($1\times1\times 0.5$ mm$^3$) using a chemical vapor deposition (CVD) technique~\cite{miyazaki2014atomistic,tahara2015quantifying,ishiwata2017perfectly,ozawa2017formation,Tsuji2022}.
The symmetry axis of the NV center (NV axis) is perfectly aligned perpendicular to the diamond surface.
The CVD-grown NV layer thickness is $2.3~\si{\micro\meter}$ [\cref{fig1}(a)], measured by secondary ion mass spectroscopy.
The areal density of NV centers and the density of nitrogen atoms are estimated to be $2.1\times10^5~\si{\micro\meter}^{-2}$ and $3\times10^{19}~\text{cm}^{-3}$, respectively.
We adhere the diamond chip to a YBCO thin film by varnish [\cref{fig1}(b)].
The stray field from the vortices is detected at the NV centers in the CVD layer at a distance $d$ away [\cref{fig1}(a)].
The YBCO sample is a (100) thin film (S-type) on a MgO substrate purchased from Ceraco Ceramic Coating GmbH.
The nominal YBCO thickness is $t_\text{sc} = 250~\si{\nano\meter}$.
The critical temperature is estimated to be $T_c=88.7~\si{K}$ from the temperature-dependent sheet resistance shown in \cref{fig1}(c).

Our microscope system is shown in \cref{fig1}(d).
The sample is fixed with vacuum grease to a stage in an optical cryostat (Montana Instruments Cryostation s50).
The sample temperature is controlled by a heater and monitored by a thermometer of the stage.
Hereafter, we use the stage thermometer value as the temperature.
We expand a green laser (532~nm, 120~mW) onto the diamond to image the photoluminescence (PL) of the NV centers.
We image the wavelength range of the NV center ($\lambda_\text{NV} = 650$--$750$~nm) with a CMOS camera and optical filters.
The optical diffraction limit is estimated to be $0.61 \frac{\lambda_\text{NV}}{\text{NA}} \sim 750$~nm.
Since we acquire the images through the diamond, the optical resolution becomes $0.9~\si{\micro\meter}$ due to optical aberration~\cite{nishimura2023tobesubmitted}.
We use a loop microwave antenna~\cite{Sasaki2016} fixed on the optical window of the cryostat to manipulate the NV centers.
A coil applies a spatially uniform static magnetic field in the direction perpendicular to the YBCO surface, parallel to the NV axis.
We perform field-cooling (FC) to generate the vortices by cooling down the stage temperature from $90~\si{K}~(>T_c)$ to the desired temperature. At the same time, we modulate their density by tuning the field generated by the coil.

Magnetic flux density is obtained using optically detected magnetic resonance (ODMR) in the NV centers~\cite{rondin2014magnetometry}.
\Cref{fig1}(e) shows typical ODMR spectra, where the vertical axis is the relative PL intensity with and without microwave irradiation, and the horizontal axis is microwave frequency.
There are two dips in each spectrum, which correspond to electron spin resonances of the NV centers.
The splitting between the resonance frequencies $\Delta f$ is larger at 1~mT (red squares) than at 0~mT (black circles), reflecting the Zeeman effect.
The splitting is given by~\cite{rondin2014magnetometry},
\begin{equation}
\Delta f = 2\sqrt{ (\gamma_{e}B_z)^2 + E^2},
\label{eq1}
\end{equation}
where $B_z$ is the magnetic flux density in the direction of the NV axis, $\gamma_e=28$~MHz/mT is the gyromagnetic ratio of an electron spin, and $E$ is a strain parameter, which is position-dependent in the crystals.
We fit the ODMR spectrum by two Lorentzian to determine $\Delta f$ and $E$ at each position and convert $\Delta f$ to $B_z$ using \cref{eq1} (described later).
Note that such a simple analysis is possible thanks to the perfectly aligned NV centers; ordinary ensemble centers have up to eight resonance signals, complicating the investigation.
In addition, the absence of the sensors oriented to other symmetric axes is beneficial for the high sensitivity because it prevents contrast reduction~\cite{ishiwata2017perfectly,tsukamoto2021vector}.
We analyze the data of whole CMOS pixels (1536~pixel $\times$ 2048~pixel) to obtain the magnetic field distribution.
To mitigate the failure of the Lorentzian fitting, we reduce shot noise by smoothing the PL image with a Gaussian filter smaller than the optical resolution (whose $1/e$ decay length is $350~\si{nm} = 5$ pixels of the camera)(see supplemantal materials for details.)

We calibrate the magnetic field of the system.
\Cref{fig1}(f) shows the dependence of the splitting $\Delta f$ on the coil voltage $V_\text{coil}$.
We obtain this data at a temperature well above $T_c$ to avoid the diamagnetism of superconductivity.
$\Delta f$ increases with increasing the absolute value of the coil voltage, as expected from Ampere's law.
The total magnetic field is obtained as,
\begin{equation}
B_z^{(T>T_c)} = \alpha_\text{coil} V_\text{coil} + B_\text{resid},
\label{eq2}
\end{equation}
where $B_z^{(T>T_c)}$ is $B_z$ at the temperature $T > T_c$, $\alpha_\text{coil}$ is the linear coefficient between coil voltage and magnetic field, and $B_\text{resid}$ is a residual magnetic field, including geomagnetism.
The solid black line in \cref{fig1}(f) is the fitting using \cref{eq1,eq2}.
It agrees well with our experimental result.
We obtain the fitting parameters 
$\alpha_\text{coil} = 1.64~\si{\micro\tesla/\milli\volt}$, and $B_\text{resid}=38.2~\si{\micro\tesla}$. 
The calibration accuracy of $B_z^{(T>T_c)}$ is $\pm1.2~\si{\micro\tesla}$ (95~\% confidence interval).

\begin{figure*}
\begin{center}
\includegraphics{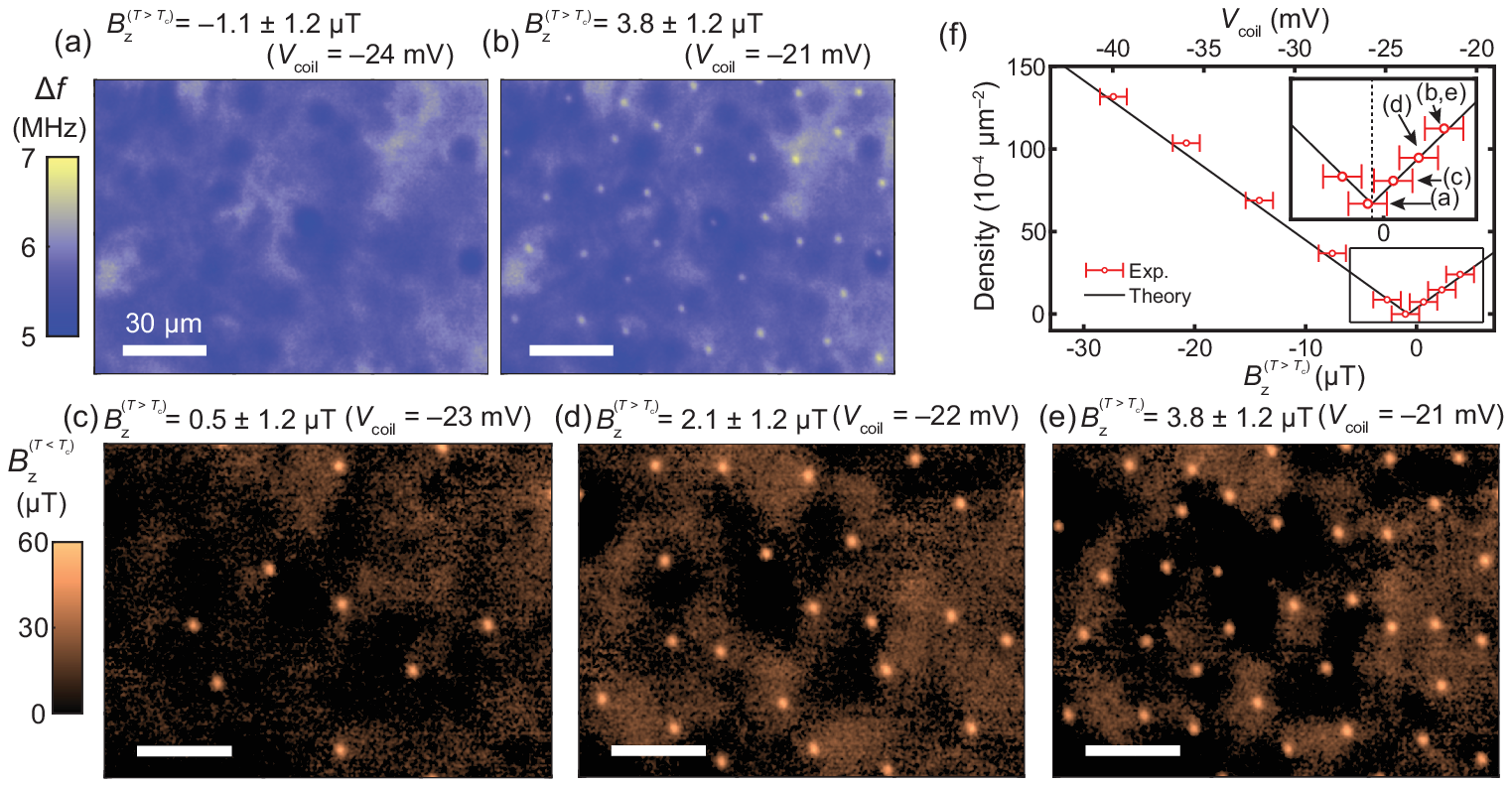}
\caption{
Magnetic imaging of superconducting vortices.
(a,b) Distribution of $\Delta f$.
(a) and (b) are the data under FC of $B_z^{(T>T_c)}=-1.1~\si{\micro\tesla}$ and $B_z^{(T>T_c)}=3.8~\si{\micro\tesla}$, respectively.
(c,d,e) Distribution of  the stray field from the vortices $B_z^{(T<T_c)}$. (see supplemantal materials for full data)
(c), (d), and (e) are the data under FC of $B_z^{(T>T_c)}=0.5~\si{\micro\tesla}$,  $2.1~\si{\micro\tesla}$, and $3.7~\si{\micro\tesla}$, respectively.
(f) Relationship between vortex density and magnetic field (coil voltage).
The error bars correspond to $\pm1.2~\si{\micro\tesla}$, the 95~\% confidence interval of the calibration [\cref{fig1}(f)].
Inset shows the enlarged view near the zero magnetic fields.
Arrows indicate the condition of each data acquired.
The measurements are performed at $40~\si{K}~(<T_c)$.
The vertical dashed black line is the actual zero field condition estimated by the linear fitting.
\label{fig2}
}\end{center}
\end{figure*}

Figures~\ref{fig2}(a) and (b) show the distributions of $\Delta f$ obtained under FC conditions of $B_z^{(T>T_c)}=-1.1~\si{\micro\tesla}$ and $B_z^{(T>T_c)}=3.8~\si{\micro\tesla}$, respectively.
We obtain these images at 40~K.
There are multiple point-shaped magnetic field distributions at the larger field [\cref{fig2}(b)], while no such distributions at the smaller field [\cref{fig2}(a)].
Each of these points is a superconducting vortex. Later we prove that they are genuinely single vortices. 
The absence of such a  feature in \cref{fig2}(a) indicates no vortices in this view, implying minuscule magnetic fields are realized in the cool-down process.
We define this condition as zero-field cooling.

Although there is no apparent vortex-like distribution in \cref{fig2}(a), there is a fluctuating distribution of $\Delta f$.
The primary cause of this phenomenon is the position-dependent strain $E$ in the diamond crystal. 
We also observe that $\Delta f$ depends on the excitation light intensity, which can lead to such a distribution~\cite{Fujiwara2020,itoh2023}(see supplemantal materials for details).
We find that the latter effect, which is smaller than that of the strain, can be efficiently removed by phenomenologically including it in strain $E$ in the following analysis. 
We calculate the magnetic field density at each pixel using
\begin{equation}
B_z = \frac{\sqrt{ \Delta f^2 - \Delta f_0^2 }}{2\gamma_{e}},
\label{eq3}
\end{equation}
where $\Delta f_0 = 2E$ is the $\Delta f$ at zero magnetic fields [\cref{fig2}(a)].
Figures~\ref{fig2}(c), (d), and (e) present the resulting magnetic field distributions $B_z^{(T<T_c)}$ obtained under FC of $B_z^{(T>T_c)}=0.5~\si{\micro\tesla}$, $2.1~\si{\micro\tesla}$, and $3.7~\si{\micro\tesla}$, respectively. 
Our analysis successfully subtracts the inhomogeneities due to the strain and excitation light intensity, and now vortices are visible more clearly.

We examine the relation between the number of vortices and the magnetic flux density during FC.
We count the number of the vortices in the field of view to obtain vortex areal density, as shown in \cref{fig2}(f).
The vortex density increases linearly with the absolute value of the magnetic field.
A superconducting vortex has a single flux quantum $\Phi_0=h/2e=2068~\si{\micro\tesla\cdot\micro\meter^2}$ (where $h$ is Planck's constant and $e$ is the elementary charge). 
The vortex density corresponds to the magnetic flux density.
Thus, in \cref{fig2}(f), the proportionality coefficient should be $\beta =4.84\times 10^{-4}~\si{\micro\meter^{-2}/\micro\tesla}$.
The solid black line is the theoretical fitting based on the calibration in \cref{fig1}(f), consistent with the experimental result within the error bars.
As shown by the vertical dashed line in the inset of \cref{fig2}(f), the zero field calibration is carried out within $-0.7~\si{\micro\tesla}$, corresponding to the exact residual field of $B_\text{resid}=37.5~\si{\micro\tesla}$, including geomagnetism. 
These results prove that the observed vortices have a single flux quantum.

\begin{figure}
\begin{center}
\includegraphics{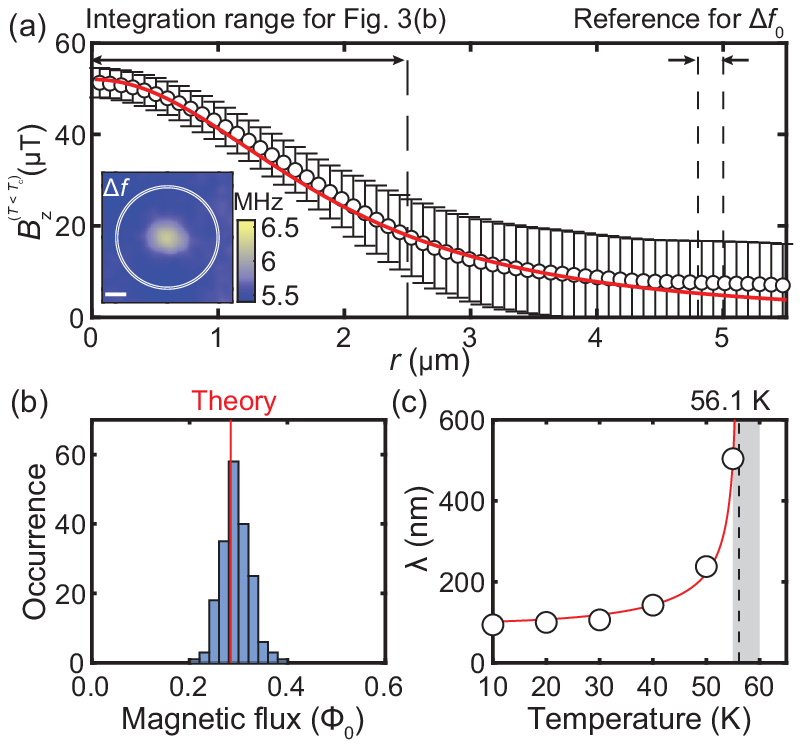}
\caption{
(a) Radial distribution of the stray field $B_z^{(T<T_c)}$ at $40\si{K}$ ($<T_c$).
The black circles and their error bar are the mean value and standard deviation for 190 vortices.
The solid red line is the theoretical fitting. 
The finite offset of the magnetic field density at $r > 4~\si{\micro\meter}$ is due to statistical errors in taking the absolute value of squared deviation\cref{eq3}.
Inset is the image of a typical single vortex in terms of $\Delta f$.
The scale bar is $2~\si{\micro\meter}$.
The circles indicate the background reference region ($4.8~\si{\micro\meter}<r<5~\si{\micro\meter}$).
(b) Histogram of the magnetic flux.
The horizontal axis shows the value obtained by integrating the magnetic flux density in a radius of $2.5~\si{\micro\meter}$ from the vortex center indicated by the arrow in the top left of \cref{fig3}(a).
The vertical red line is the value obtained from the theoretical treatment (see text).
(c) Temperature dependence of the London penetration depth $\lambda$.
The gray area indicates that the local temperature of the measurement position reaches the critical temperature due to laser irradiation.
\label{fig3}
}\end{center}
\end{figure}

The present method, which observes many vortices in a wide field of view quantitatively and simultaneously, enables us to make a statistical analysis. 
The inset of \cref{fig3}(a) depicts the distribution of $\Delta f$ for a typical vortex.
Thus, the magnetic field is isotropically distributed concerning the distance $r$ from the vortex center.
We rely on \cref{eq3} to extract the field, where we define $\Delta f_0$ as an average of $\Delta f$ far away from the vortex center (specifically, $4.8~\si{\micro\meter} < r < 5.0~\si{\micro\meter}$) to avoid the effect of drift during FC cycles.
There are 290 vortices in the results obtained under FC of several $B_z^{(T>T_c)}$ between $-13.9~\si{\micro\tesla}$ and $5.3~\si{\micro\tesla}$.
We estimate the center-of-mass positions of these vortices by Gaussian fitting. 
Among them, we extract 190 vortices, located away from large inhomogeneity and separated by more than 8~$\si{\micro\meter}$ to avoid the effect of drift and the influence of stray fields from neighboring vortices.

\Cref{fig3}(a) shows the obtained distribution of the magnetic field of a vortex as a function of $r$.
The error bar reflects the standard deviation concerning the 190 vortices used in the analysis.
The magnetic field just above the vortex center is $51.1~\si{\micro\tesla}$, while the error bars are kept as small as $\pm 5.47~\si{\micro\tesla}$.

\Cref{fig3}(b) shows the magnetic flux projection obtained by integrating each vortex field over the region of $r<2.5~\si{\micro\meter}$, as indicated by the arrow in the top left of \cref{fig3}(a). 
The histogram forms a Gaussian distribution, meaning that all the single vortices are accurately captured as having the same flux.
The magnetic flux's average and standard deviation is $0.295~\Phi_0$ and $0.029~\Phi_0$, respectively, showing that the present technique has a precision of 10~\%.
The statistical uniformity also guarantees that our analysis has successfully removed the observed inhomogeneities. $0.295~\Phi_0$ is smaller than $\Phi_0$ because the integration range is limited to $r<2.5~\si{\micro\meter}$ and 
only the field component parallel to the NV axis is detected, as schematically shown in  \cref{fig1}(a).

Next, we quantitatively compare the distribution of the stray field with theory. 
The stray field from a quantum vortex exhibits different characteristic lengths in bulk \cite{clem1975, carneiro2000vortex, kogan2003} 
and thin-film\cite{carneiro2000vortex, kogan2003, kogan2021} cases, dictated by the London penetration depth $\lambda$ and the Pearl length\cite{Pearl1964} $\Lambda = 2\lambda^2/t_\text{sc}$, respectively. 
Given that the thickness is $t_\text{sc}\sim 250$~nm in our case, comparable to the penetration length $\lambda$ (a few hundred nm \cite{Djordjevic2002,Sonier2007,abou2021magnetic}), 
we analyze our results using the model derived from \citet{carneiro2000vortex}, which is applicable to both bulk and thin-film cases(see supplemantal materials for details):
\begin{align}
  \small
B_z^{(T<T_c)}(r;d,\lambda) = \frac{\Phi_0}{2\pi\lambda^2} \int_0^{\infty} dk \frac{kJ_0(kr)}{k^2+\lambda^{-2}} f(k,d), \notag\\
f(k,d)=\frac{(k+\tau) e^{\tau t_\text{sc}} + (k-\tau) e^{-\tau t_\text{sc}} -2k}{(k+\tau)^2 e^{\tau t_\text{sc}} - (k-\tau)^2 e^{-\tau t_\text{sc}}} \tau e^{-kd},
\label{eq4}
  \normalsize
\end{align}
where $J_0$ is 0-th order Bessel function of the first kind, $\tau = \sqrt{k^2 + \lambda^{-2}}$, and $\lambda$ is the London penetration depth, which depends on temperature.
Our method is subject to the influence of the thickness of the CVD layer and the optical resolution.
The solid red line in \cref{fig3}(a) results from the fitting using a spatially integrated form 
of \cref{eq4} to include these effects, reproducing the experimental result well within the error bars.
The calculated flux is also consistent with the statistical results of the magnetic flux shown by the red vertical line in \cref{fig3}(b). 
We obtain $\lambda = 154~\si{nm}$ when we fix $d = 1.35~\si{\micro\meter}$. 
Since we repeat thermal cycles several times and confirm that two-parameter estimation from fitting both $d$ and $\lambda$ always yields a value of $d$ around $1.35~\si{\micro\meter}$, we fix $d = 1.35~\si{\micro\meter}$ hereafter.
The vortex size in a superconducting thin film, i.e., the Pearl length, is estimated to be $\Lambda=190$~nm, smaller than the optical resolution.
The stray field distribution from the vortex appears larger than $\Lambda$ because the sensor ensemble is located away by $d$ from the YBCO film, which disperses the magnetic flux, as shown in \cref{fig1}(a).

We investigate the temperature dependence of $\lambda$.
\Cref{fig3}(c) shows the $\lambda(T)$ from fitting the experimental result $B_z^{(T<T_c)}(r;d,\lambda)$ at each temperature obtained by raising temperature after FC of $B^{T>T_c} = -20.8~\si{\micro T}$.(see supplemantal materials for full data)
The resulting $\lambda (T)$ remains at $\sim 100$~nm from $10~\si{K}$ to $30~\si{K}$ but dramatically increases above $40~\si{K}$, reaching $\sim 500~\si{nm}$ at $55~\si{K}$.
The vortex disappears at $T'_c$ between $55~\si{K}$ and $60~\si{K}$ [a gray area in \cref{fig3}(c)], lower than the original $T_c=88.7$~K, due to the local heating by laser irradiation. 

Previous studies report that $\lambda$ varies from a minimum of 130~nm to a maximum of 810~nm~\cite{Djordjevic2002,Sonier2007,abou2021magnetic}. 
The observed behavior of $\lambda(T)$ is consistent with them. 
We fit the temperature dependence of $\lambda$ using the following empirical model for a $d$-wave superconductor~\cite{prohammer1991,basov2005,stilp2014},
\begin{equation}
\lambda(T) = \frac{\lambda(0)}{\sqrt{1-(T/T'_c)^2}}.
\label{eq5}
\end{equation}
We obtain $\lambda(0) = 100~\si{nm}$ and $T'_c = 56.1~\si{K}$; the fitted curve agrees well with the obtained $\lambda(T)$.
In some models \cite{Pearl1964,auslaender2009,acosta2019color}, the covariance of $d$ and $\lambda$ is large, meaning that $\lambda$  might vary depending on $d$ (and vise versa), and they might not be well determined by two-parameter fitting.
Nevertheless, estimating the scaling behavior of one parameter from the fitting with the other parameter fixed is still meaningful in such a situation. 
The penetration depth $\lambda$ is an essential phenomenological parameter in describing superconductivity, and various methods have studied its behavior. 
Although the present method is not immune from the effect of laser heating, it provides an important alternative to systematically address this parameter under a wide range of experimental conditions. 

To conclude, we have quantitatively established the wide-field imaging of superconducting vortices using a perfectly aligned diamond quantum sensor.  
By eliminating the effect of inhomogeneity, the magnetic flux of a single vortex in a YBCO thin film was visualized with an accuracy of $\pm 10$~\%. 
In addition, we demonstrate the quantitative method to examine the penetration depth.
We can further improve sensitivity and accuracy by combining techniques such as multi-frequency magnetic resonance~\cite{kazi2021wide} and thinner CVD layers~\cite{ishiwata2017perfectly}.
The demonstrated precise high throughput method, applicable over a wide temperature range, helps explore various superconducting properties and statistical evaluation, including their MHz - GHz dynamics~\cite{degen2017quantum}.
For example, it could apply to investigating an anomalous quantum vortex, such as a half-integer one, and to the high-pressure superconductivity in diamond anvil cells~\cite{Hsieh2019,Lesik2019,Toraille2020}.

See the supplemental materials for all the magnetic imaging data in the present experiment, details of the numerics employed for the analysis, descriptions of the fitting methods, and information regarding the sensitivity.

We appreciate K. M. Itoh (Keio University) for providing the cryostat.
The authors acknowledge the support of Grant-in-Aid for Scientific Research (Nos.~JP22K03524, JP19H00656, JP19H05826, and JP22H04962) and of the MEXT Quantum Leap Flagship Program (Grant No.JPMXS0118067395). 
Some parts of this work were conducted at (Takeda Clean Room, Univ. Tokyo and Nanofab, Tokyo Tech), supported by Advanced Research Infrastructure for Materials and Nanotechnology in Japan (ARIM), Grant Number JPMXP1222UT1131 and JPMXP1222IT0058. 
SN is supported by the Forefront Physics and Mathematics Program to Drive Transformation (FoPM), WINGS Program, The University of Tokyo, and JSR fellowship.

\pagebreak
\widetext
\setcounter{section}{0}
\setcounter{equation}{0}
\setcounter{figure}{0}
\setcounter{table}{0}
\setcounter{page}{1}
\makeatletter
\renewcommand{\theequation}{S\arabic{equation}}
\renewcommand{\thefigure}{S\arabic{figure}}
\renewcommand{\bibnumfmt}[1]{[S#1]}
\renewcommand{\citenumfont}[1]{S#1}

\section*{Supplemental Information for:\\
Wide-field quantitative magnetic imaging of superconducting vortices \\
using perfectly aligned quantum sensors
}

This Supplemental Material is organized as follows:
\Cref{sec:process_ODMR} provides additional information related to the processing of pixel-wise ODMR.
\Cref{sec:opt_res} describes the evaluation of optical resolution, including the effects of smoothing.
\Cref{sec:full_data} shows all the results obtained in this measurement with varying magnetic fields during field cooling (FC).
\Cref{sec:model} describes the details of the numerics of the theoretical model and fitting procedures by this model.
\Cref{sec:SensorSensitivity} details the sensor sensitivity.

\section{Processing of ODMR Spectra}
\label{sec:process_ODMR}

\subsection{Strain parameter and optical power intensity}
In the main text, we attribute the finite splitting of resonance frequency $\Delta f$ measured in zero magnetic fields mainly to a local strain distribution. 
By calibrating $\Delta f$ relying on Eq.~(1) in the main text, we have successfully removed the inhomogeneities. 

\begin{figure}[tbp]
  \begin{center}
  \includegraphics[width = 0.9\hsize]{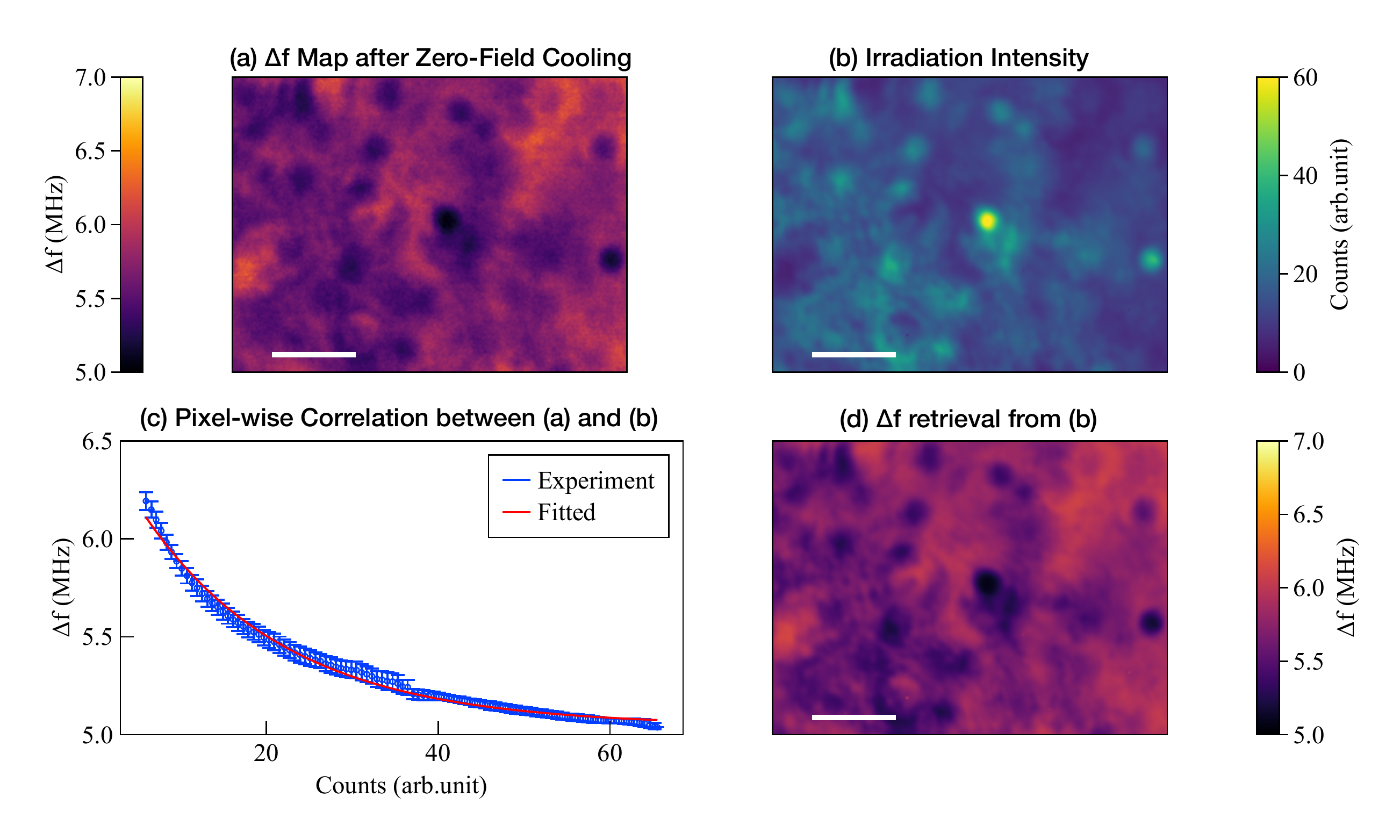}
  \caption{
    Comparison between $\Delta f$ distribution and intensity distribution of excitation light under near zero fields (results of zero-field cooling down to $40~\si{K}$).
    (a) $\Delta f $ distribution in zero-field at $40~\si{K}$. This figure is the reproduction of Fig 2(a) in the main text.
    (b) PL intensity distribution obtained for reference without microwave application in the measurement in \cref{fig:opt-df}(a)
    (c) Pixel-wise correlation of PL intensity (\cref{fig:opt-df} (b)) to $\Delta f$ (\cref{fig:opt-df} (a)). 
    (d) $\Delta f$ distribution deduced from the results of \cref{fig:opt-df} (b) and the exponential fitting curve in \cref{fig:opt-df} (c).
  \label{fig:opt-df}
  }\end{center}
\end{figure}

In the main text, we briefly mention that $\Delta f$ also depends on the excitation light intensity, which can lead to similar fluctuation. 
We have found that the distribution of $\Delta f$ at zero fields depends on the intensity of excitation light, although this effect is much smaller than that of the strain.
\Cref{fig:opt-df}(a) replicates Fig.~2(a) in the main text, which shows the intensity plot of $\Delta f$ after the zero-field cooling down to $40~\si{K}$.
There are no superconducting vortices, which means that average field flux density $B_\text{avg}$ is extremely low (at least $\vert B_\text{avg}\vert < 140~\si{nT}$).
However, there exist complex $\Delta f$ inhomogeneities.  
\Cref{fig:opt-df}(b) shows the optical intensity during this measurement without  the microwave being applied. 
The fluctuating patterns in Figs.~1(a) and 1(b) are apparently in a negative correlation.

The fact that $\Delta f$ depends on the optical intensity was reported for the nano-diamond NV center ensemble~\cite{si_Fujiwara2020}.
We have recently confirmed that a similar phenomenon occurs in a bulk diamond crystal~\cite{itoh2023}.
The present experiment using (111)-oriented bulk diamond also exhibits such a dependence. 

We further investigate the optical intensity dependence. 
\Cref{fig:opt-df} (c) shows the correlation between photoluminescence (PL) counts and $\Delta f$ for each pixel.
Here the PL count is rolled into 101 bins, and the scatter plot indicates the mean value of $\Delta f$ when PL takes a certain value inside the bin.
The error bar shows the standard deviation of $\Delta f$.
$\Delta f$ shows an exponential decay against the PL intensity. 
The solid red line is an exponential fit, reproducing well the decaying behavior.
Based on this fitting, we can mitigate the effect of $\Delta f$ due to the optical intensity. 
\Cref{fig:opt-df} (d) shows the $\Delta f$ distribution calculated from the solid red curve in \cref{fig:opt-df} (c) and the PL intensity distribution shown in \cref{fig:opt-df} (b). 
\Cref{fig:opt-df} (d) reproduces \cref{fig:opt-df} (a) well: the root-mean-squared (RMS) deviation between Figs. 1(a) and (d) is $0.051~\si{MHz} \sim 1 \%$. 
The splitting in resonance frequency under near zero field is further discussed in \cite{itoh2023}, where we propose possible mechanisms.

\subsection{Pixel-wise ODMR spectrum fitting}
In the main text, we perform ODMR measurements in a wide field of view, and we obtain the magnetic flux density by fitting the ODMR spectrum on each pixel of the CMOS camera.
The fitting calculations for all $2048 \times 1536$ pixels of the CMOS censor are executed in parallel 
using distributed memory multi-process computing implemented in Julia language~\cite{julia}. 
It typically takes three minutes to compute using a standard commercial computer.

\subsection{Image Smoothing}
\label{subsec:smoothing}
We apply an image-smoothing technique to improve the signal-to-noise ratio (SNR)~\citep{haddad}.
This method interprets the acquired ODMR spectra per pixel as a bundle of images for each applied microwave frequency.
Each image of a given frequency is blurred by the optical resolution. 
Therefore, it is possible to perform image smoothing to the same or less extent as the scale of the optical resolution.

Specifically, we apply Gaussian convolution by a Gaussian kernel whose $1/e$-width is $350~\si{\nano\meter}$.
The convolution is expressed as follows: 
\begin{equation}
 C_{i,j} = \sum_{kl} C_{i+k,j+l} K_{k,l}.
 \label{eq:conv}
\end{equation}
Here, we denote the PL intensity at the $i,j$-th pixel as $C_{ij}$ and the kernel function as $K_{k,l}$. 
We adopt the following Gaussian kernel that takes $~\sigma_G$ as the $1/e$ width,
\begin{equation}
 K_{k,l} = G_{k,l}(\sigma_G) = \exp[-\frac{k^2 + l^2}{\sigma_G^2}].
\end{equation}
\Cref{eq:conv} represents the addition of the counts of surrounding pixels weighted relative to the magnitude of the counts at $i,j$-th pixel, $C_i,j$, set to 1. This sort of convolution increases the total amount of PL counts for each pixel compared to that without the convolution is applied. 
The PL counts with the convolution is multiplied by a factor of,
\begin{equation}
\iint \exp[-\frac{k^2 + l^2}{\sigma_G^2}] dS = \pi\sigma_G^2.
\end{equation}
Thus, the SNR of the spectrum improves by about the magnitude of the squared root value $\sqrt{\pi}\sigma_G$.

\section{Optical resolution}
\label{sec:opt_res}
We evaluate the effective optical resolution of our system. 
Hereafter, we use $1/e$ radii to indicate it.

\subsection{Effect of optical aberration}
The experimental optical resolution is subject to optical aberration~\cite{born_wolf,si_nishimura2023tobesubmitted}.
Specifically, we assume the $1/e$ radius
\begin{equation}
 \sigma_\text{opt} \sim 610~\si{\nano\meter}
 \label{eq:sigma_opt}
\end{equation}
for an optical system where the numerical aperture (NA) of the objective lens is $0.55$, and the measured PL wavelength $\lambda$ is $700~\si{\nano\meter}$.
This assumption is based on calculating the point spread function of a single NV center when observed through diamonds with a thickness of around 500~$\si{\micro m}$.
In addition to optical aberration, the optical resolution is degraded due to image smoothing, described in the following subsection.

\subsection{Effect of image smoothing}\label{subsec6:analysis}
We evaluate the optical resolution loss due to image smoothing by Gaussian convolution.
The actual resolution $\sigma_f$ is given as
\begin{equation}
 \sigma_f = \sqrt{\sigma_\text{opt}^2 + \sigma_G ^2} \sim 780~\si{\nano\meter},
 \label{eq:sigma_f}
\end{equation}
where $\sigma_\text{opt}=610~\si{nm}$ is the raw optical resolution given in \cref{eq:sigma_opt}, and $~\sigma_G=345~\si{nm}$ is the $1/e$ decay length of the Gaussian kernel.
Assuming that the point spread function representing NV centers' PL intensity distribution can be approximated as a Gaussian function, \cref{eq:sigma_f} is derived as follows.
The Gaussian distribution $G(r,\sigma)$ is defined as
\begin{equation}
 G(r,\sigma) = \exp(-\frac{r^2}{\sigma^2}).
\end{equation}
Fourier transformation of this kernel is, 
\begin{equation}
 \mathcal{F}[G](\xi,\sigma) = \frac{1}{\sqrt{2}\sigma}\exp(-\frac{\sigma^2 r^2}{4}).
\end{equation}
Therefore, the convolution of two Gaussian functions $G_1$ and $G_2$ with variances of $\sigma_1$ and $\sigma_2$, respectively, is given by the Fourier transformed form as, 
\begin{align}
 \mathcal{F}[G_1 *G_2](\xi) 
 &= \mathcal{F}[G_1](\xi)\mathcal{F}[G_2](\xi)\notag\\
 &= \frac{1}{2\sigma_1\sigma_2}\exp(-\frac{(\sigma_1^2 +\sigma_2^2)r^2}{4}).
\end{align}
Thus, we have 
\begin{equation}
 G_1 *G_2(r) = \frac{\sqrt{\sigma_1^2 +\sigma_2^2}}{\sqrt{2}\sigma_1\sigma_2}
 \exp(-\frac{r^2}{\sigma_1^2 +\sigma_2^2})\propto G(r, \sqrt{\sigma_1^2 + \sigma_2 ^2}).
\end{equation}

\section{Full data obtained with varying flux density in Field Cooling}
\label{sec:full_data}
We give all the data used for Fig.~2(f) in the main text here.
The results of magnetic field distribution $B_z^{(T<T_c)}$ at 
40~K for various FC conditions of $B_z^{(T>T_c)}$, are shown in \cref{fig:Bmap_fine}.
Furthermore, the results of magnetic field distribution $B_z^{(T<T_c)}$ obtained by raising temperature after cooldown in a field of $B_z^{(T>T_c)} = -20.8~\si{\micro T}$, are shown in Fig. S3.

\begin{figure}[p]
 \begin{center}
 \includegraphics[width = 0.85\hsize]{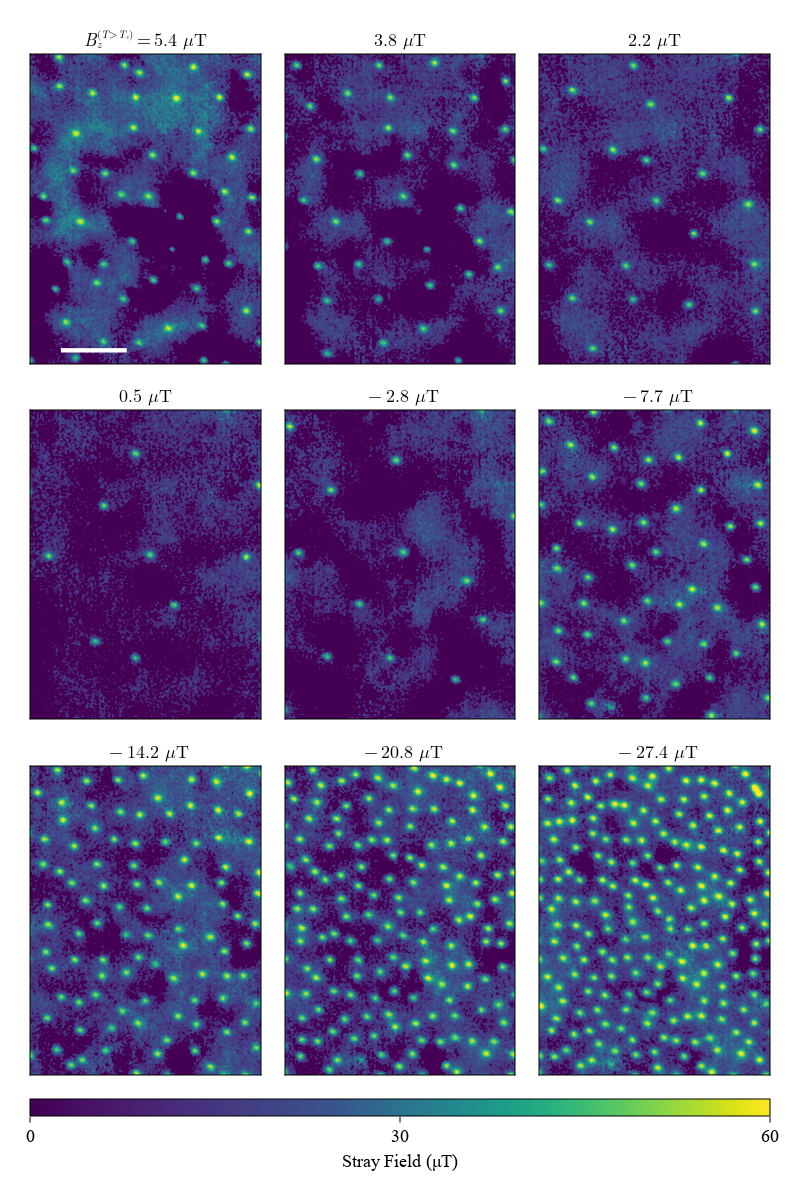}
 \caption{
 Magnetic field distribution $B_z^{(T<T_c)}$ under FC of various $B_z^{(T>T_c)}$. Scalebar is 30~\si{\micro m}.
 \label{fig:Bmap_fine}
 }\end{center}
\end{figure}

\begin{figure}[h]
  \begin{center}
  \includegraphics[width = 0.85\hsize]{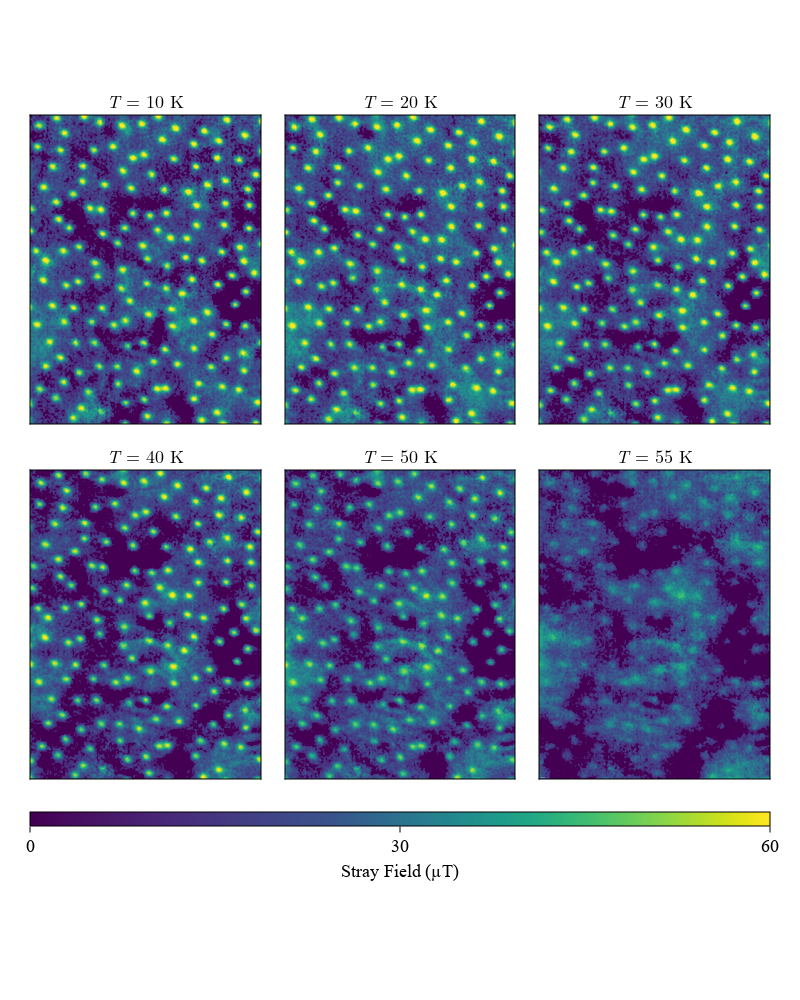}
  \caption{
  Magnetic field distribution $B_z^{(T<T_c)}$ obtained at various temperature.  Scalebar is 30~\si{\micro m}.
  \label{fig:Bmap_fine}
  }\end{center}
 \end{figure}

\newpage

\section{Numerics of the theoretical model and fitting procedures}
\label{sec:model}
This section gives the procedure to calculate the theoretical model of vortex stray field distribution \cite{si_carneiro2000vortex}.

\subsection{Numerics of the theoretical model}
The stray field $B_r(r,z,\lambda)$ and $B_z(r,z,\lambda)$ from a vortex are given as functions of $r$ and $z$ using the London penetration depth $\lambda$ in the cylindrical geometry as [Eq.~(8) of Ref.~\cite{si_carneiro2000vortex}]:
\begin{align}
  B_r(r,z,\lambda) &= \frac{\Phi_0}{2\pi \lambda^2} \int_0^\infty dk \frac{J_1(kr) }{k^2 + \lambda^{-2}}g(k,z)\\
  B_z(r,z,\lambda) &= \frac{\Phi_0}{2\pi \lambda^2} \int_0^\infty dk \frac{kJ_0(kr) }{k^2 + \lambda^{-2}}f(k,z).\label{eq6:Bz}
\end{align}
Here, the $J_i(x)$ is the first-kind Bessel function of $i$-th order. Also, 
\begin{equation} 
  f(k,z) = 
  \begin{cases}
    c_1 e^{-kz} &z>0\\
    1 + c_2 e^{\tau z } + c_3 e^{-\tau z }& -d \leq z \leq 0\\
    c_1 e^{k(z+d)} &z<-d
  \end{cases},
\end{equation}
and
\begin{equation}
  g(k,z) = -{\pdv{f}{z}}  {(k,z)}, 
\end{equation}
with
\begin{equation}
  \tau = \sqrt{k^2 + \lambda^{-2}}.
\end{equation}
The coefficients $c_i~(i=1, 2, 3)$ are expressed as, 
\begin{align}
  c_1(k) &= \frac{(k+\tau) e^{\tau d} + (k-\tau) e^{-\tau d} -2k}{(k+\tau)^2 e^{\tau d} - (k-\tau)^2 e^{-\tau d}} \tau\notag,\\
  c_2(k) &= \frac{(k+\tau) e^{\tau d} + (k-\tau) }{(k+\tau)^2 e^{\tau d} - (k-\tau)^2 e^{-\tau d}}k\notag,\\
  c_3(k) &= \frac{(k-\tau) e^{-\tau d} - (k+\tau) }{(k+\tau)^2 e^{\tau d} - (k-\tau)^2 e^{-\tau d}}k .
\end{align}

This model, although not explicitly incorporating the so-called Pearl length $\Lambda = 2\lambda^2/d$, is applicable to both bulk and thin film cases, as stated in [Ref. \cite{si_carneiro2000vortex}].
Specifically, in the case of thin films, it can be simplified into the formula \cite{si_kogan2003,si_kogan2021} that explicitly includes $\Lambda$:
\begin{align}
  B_z^{(T<T_c)}(r;d,\lambda) &= \frac{\Phi_0}{2\pi\lambda^2} \int_0^{\infty} dk \frac{kJ_0(kr)}{k^2+\lambda^{-2}} f(k,d) \tag{[\text{Eq. (4)}]}\\
  &= \frac{\Phi_0}{4\pi^2}\iint d^2k \frac{e^{ik(r-z)}}{1+(k\lambda)^2}f(k,d) \notag\\
  &= \frac{\Phi_0}{4\pi^2}\iint d^2k \frac{e^{ik(r-z)}}{1+(k\lambda)^2}\frac{(k+\tau) e^{\tau d} + (k-\tau) e^{-\tau d} -2k}{(k+\tau)^2 e^{\tau d} - (k-\tau)^2 e^{-\tau d}} \tau \tag{[Eq. (A12,A15) in SI]}\\
  &\approx \frac{\Phi_0}{4\pi^2}\iint d^2k \frac{e^{ik(r-z)}}{1+(k\lambda)^2}
  \frac{(k+\tau) (1+\tau d) + (k-\tau)  (1-\tau d) -2k}{(k+\tau)^2  (1+\tau d) - (k-\tau)^2  (1-\tau d)} \tau \tag{$e^{\pm \tau d}\sim 1\pm \tau d$ when $d/\lambda \ll 1$}\\
  &=  \frac{\Phi_0}{4\pi^2}\iint d^2k \frac{e^{ik(r-z)}}{1 + 2k\lambda^2(1+{kd})/d}\notag\\
  &= \frac{\Phi_0}{4\pi^2}\iint d^2k \frac{e^{ik(r-z)}}{1+k\Lambda(1+kd)}\notag\\
  &\approx \frac{\Phi_0}{4\pi^2}\iint d^2k \frac{e^{ik(r-z)}}{1+k\Lambda}
  \tag{Eq. (9) in  Ref. \cite{si_kogan2021}}.
\end{align}

In the main text, we calculate \cref{eq6:Bz}, which is expressed in the Fourier-integral form,  using adaptive-step numerical integration \cite{davis2007} with a finite cut-off of wavenumber.
Next, we calculate the following integral $B_\text{exp}(r; d,\lambda)$ to simulate the experimental result, taking into account the optical resolution and the CVD-grown NV layer thickness ($t_\text{NV}$):
\begin{align}
  B_\text{exp}(r; d,\lambda) &= \int_d^{d+t_\text{NV}} B_z(r,z,\lambda) * K(r;\sigma_f,z) dz
\end{align}
Here, we assume the depth of focus is sufficiently large, which means the kernel of the optical resolution $K(r;\sigma_f,z)$ does not depend on the $z$-coordinate. Thus we first execute integral along depth-wise of the NV center layer. 
Then, we calculate the convolution, 
\[
  \qty[ \int_d^{d+t_\text{NV}} B_z(r,z,\lambda)dz] * K(r;\sigma_f), 
\]
where $K(r)$ is the approximated Gaussian kernel corresponding to the optical resolution.
\subsection{Fitting procedures}
For the fitting, we minimize the mean squared error (MSE) of the experimental results and the function defined above.
This model includes multiple processes of numerical integration and is time-consuming. 
We thus explore the minimum of MSE by Gradient-less search.
In the present result, we start with the 10K data obtained for Fig. 3(c), 
and  we conducted two-parameter fitting as following steps; we first fix $d$ and optimize $\lambda$
by bounded univariate optimization (Brent's method ~\citep{brent}). 
The optimal values for $d$ and $\lambda$ are then determined by Brute force varying $d$ in 0.01 increments.
As described in the main text, this model is not highly reliable~\cite{transtrum2011}: it is confirmed that if changing $d$ by about 5\% ($1.3\lesssim d\lesssim 1.4$) 
and setting the optimal value of $\lambda$ for $d$, the MSE value typically changes by less than only 1\%, implying that the best fit might not be so meaningful.
However, we also confirmed that changing $\lambda$ by 5\% with a fixed $d$ (or vice versa) increases the MSE by typically about 20\%. 
Thus, fitting $\lambda$ with a fixed $d$ is still considered to yield more robust information about $\lambda$ scaling.

\section{Sensor Sensitivity}
\label{sec:SensorSensitivity}
We evaluate the sensitivity of our sensor as follows.

First, we determine the confidence interval of the flux density obtained from fitting the ODMR spectra. 
An ODMR spectrum consists of Lorentzian forms, 
\begin{equation}
  L(x; \beta_1, \beta_2, \beta_3) = \frac{\beta_3}{(x - \beta_1) ^2 + (\beta_2)^2},
\end{equation}
where $\beta_1$ corresponds to the resonance frequency, 
$\beta_2$ to the resonance linewidth, and $\beta_3$ to the contrast. 
The model function of an ODMR spectrum is, 
\begin{equation}
  f(x; D, B, E, \vec{\beta}) = 1- L(x, D-\sqrt{(\gamma_e B)^2 + E^2}, \beta_1, \beta_2) - L(x, D+\sqrt{(\gamma_e B)^2 + E^2}, \beta_1, \beta_2). \label{eq6:f}
\end{equation}
Here, $B$ is the magnetic field, $E$ is the strain parameter, and $D$ is the zero field splitting. 
Note that $E$ is pre-determined to a fixed value $\Delta f_0$.
Consider the case of fitting an experimental ODMR spectrum $\vec{y}$ using this model function. 
The Jacobian $J$ and the residual vector $\vec{r}$ are expressed as,
\begin{equation}
  J = (J_{ij})  =\pdv{f(x;\vec{\beta})}{\beta_j} ,\quad \vec{r} = (r_i) = y_i -f (x_i).
\end{equation}
The index $i$ corresponds to each component of the microwave frequency, 
and $j$ specifies the fitting parameters. We calculate the covariance matrix, 
\begin{equation}
  \mathrm{Cov} = (J^\intercal J)^{-1} \|r\|^2, \label{eq6:cov}
\end{equation}
to obtain the confidence interval, which is defined as the root value of the diagonal components of the covariance matrix~\cref{eq6:cov}, multiplied by 
Student's $t$ distribution~\citep{hansen_lsq}. 

Next, we evaluate the sensitivity from the obtained confidence interval of the magnetic flux density.
We perform fitting using the data with the longest integration time and fix the model function with the parameters obtained at this time. 
Then, we calculate $r_i(T_\mathrm{int})$ for each pair of data $y_i(T_\mathrm{int})$ with varying integration time $T_\mathrm{int}$. 
The decay rate determines the sensitivity to magnetic flux density $\eta$ in proportion to the square root of the integration time $T_\mathrm{int}$ according to shot noise. 
Precisely, we determine $\eta$ by using the following model,
\begin{equation}
  \Delta B(T_\mathrm{int}) = \frac{\eta}{\sqrt{T_\mathrm{int}}} + \Delta B_0(T_\mathrm{int}).
\end{equation}
Here $\eta$ is determined for each pixel of the camera. Due to the nonlinearity of the model function $f$ [\cref{eq6:f}] concerning the magnetic field, the sensitivity is also spatially distributed according to the magnetic flux density distribution.
Such a distribution is averaged as follows,
\begin{equation}
  \Delta B_\mathrm{avg} = \qty(\mathrm{avg} \qty(\frac{1}{\Delta B ^2}))^{-1/2}, 
  \label{eq6:sens_scale}
\end{equation}
where $\mathrm{avg}$ means pixel-wise average, meaning the inverse of the time to achieve a specific variance on average.
\begin{figure}[tbp]
  \begin{center}
  \includegraphics[width = 0.8\hsize]{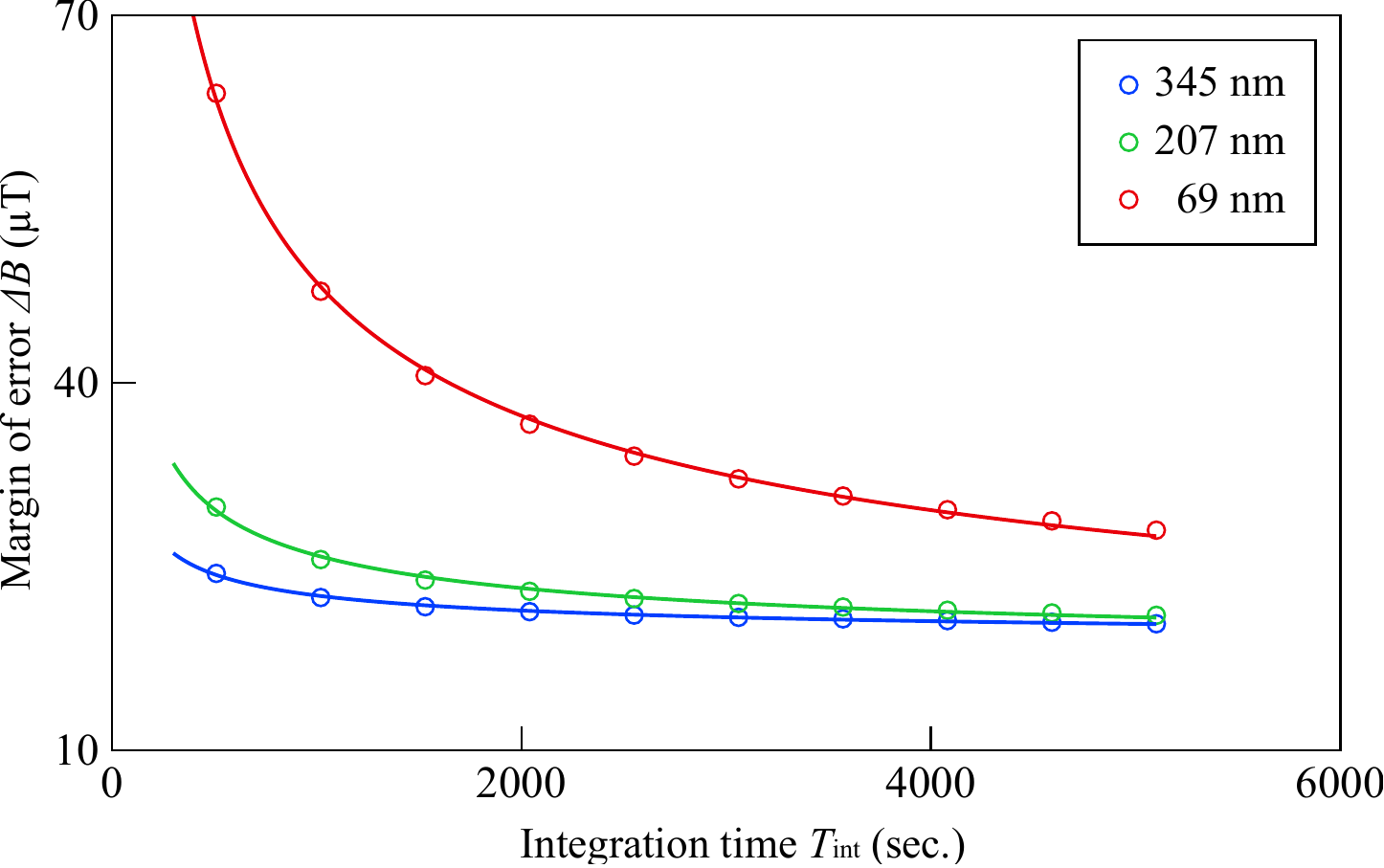}
  \caption{Scaling of $\Delta B$ over $T_\text{int}$ (the sensitivity $\eta$) with varying kernel size of convolution.
  \label{fig:sens}
  }\end{center}
\end{figure}

\Cref{fig:sens} shows the resulting dependence of $\Delta B$ on the integration time for each size of the convolution range. 
The legends represent the $1/e$ radii of the Gaussian filter. 
Both results show squared root decay [\cref{eq6:sens_scale}]. 
The respective solid lines represent the fitting. 
We obtain 
\begin{equation}
  \eta = 
  \begin{cases} 
    132~\si{\micro\tesla/\sqrt{\hertz }} & \text{for} \, ~\sigma_G = 5\times\text{pixels} = 345~\si{\nano\meter}\\
    288~\si{\micro\tesla/\sqrt{\hertz }} & \text{for} \, ~\sigma_G = 3\times\text{pixels} = 207~\si{\nano\meter}\\
    1180~\si{\micro\tesla/\sqrt{\hertz }} & \text{for} \, ~\sigma_G = 1\times\text{pixel} = 69~\si{\nano\meter}
  \end{cases}.
\end{equation}
The sensitivity is consistent with the results shown in \citep{si_tsukamoto2021vector}, and is improved by increasing $\sigma_G$.
The gain in SNR by calculating convolution is $\sqrt{\pi} ~\sigma_G$, as shown in~\Cref{subsec:smoothing}. 
The sensitivity evaluation does not obey a simple linear scaling. 
The sensitivity has a distribution depending on the magnetic field, and the optimal fitting result should depend on the kernel size $\sigma_G$. 
Such an effect is not considered in calculating the margin of error, which possibly explains the absence of linear scaling.


\begin{thebibliography}{53}%
\makeatletter
\providecommand \@ifxundefined [1]{%
 \@ifx{#1\undefined}
}%
\providecommand \@ifnum [1]{%
 \ifnum #1\expandafter \@firstoftwo
 \else \expandafter \@secondoftwo
 \fi
}%
\providecommand \@ifx [1]{%
 \ifx #1\expandafter \@firstoftwo
 \else \expandafter \@secondoftwo
 \fi
}%
\providecommand \natexlab [1]{#1}%
\providecommand \enquote  [1]{``#1''}%
\providecommand \bibnamefont  [1]{#1}%
\providecommand \bibfnamefont [1]{#1}%
\providecommand \citenamefont [1]{#1}%
\providecommand \href@noop [0]{\@secondoftwo}%
\providecommand \href [0]{\begingroup \@sanitize@url \@href}%
\providecommand \@href[1]{\@@startlink{#1}\@@href}%
\providecommand \@@href[1]{\endgroup#1\@@endlink}%
\providecommand \@sanitize@url [0]{\catcode `\\12\catcode `\$12\catcode
  `\&12\catcode `\#12\catcode `\^12\catcode `\_12\catcode `\%12\relax}%
\providecommand \@@startlink[1]{}%
\providecommand \@@endlink[0]{}%
\providecommand \url  [0]{\begingroup\@sanitize@url \@url }%
\providecommand \@url [1]{\endgroup\@href {#1}{\urlprefix }}%
\providecommand \urlprefix  [0]{URL }%
\providecommand \Eprint [0]{\href }%
\providecommand \doibase [0]{https://doi.org/}%
\providecommand \selectlanguage [0]{\@gobble}%
\providecommand \bibinfo  [0]{\@secondoftwo}%
\providecommand \bibfield  [0]{\@secondoftwo}%
\providecommand \translation [1]{[#1]}%
\providecommand \BibitemOpen [0]{}%
\providecommand \bibitemStop [0]{}%
\providecommand \bibitemNoStop [0]{.\EOS\space}%
\providecommand \EOS [0]{\spacefactor3000\relax}%
\providecommand \BibitemShut  [1]{\csname bibitem#1\endcsname}%
\let\auto@bib@innerbib\@empty
\bibitem [{\citenamefont {Fisher}\ \emph {et~al.}(1991)\citenamefont {Fisher},
  \citenamefont {Fisher},\ and\ \citenamefont {Huse}}]{fisher1991thermal}%
  \BibitemOpen
  \bibfield  {author} {\bibinfo {author} {\bibfnamefont {D.~S.}\ \bibnamefont
  {Fisher}}, \bibinfo {author} {\bibfnamefont {M.~P.~A.}\ \bibnamefont
  {Fisher}},\ and\ \bibinfo {author} {\bibfnamefont {D.~A.}\ \bibnamefont
  {Huse}},\ }\href {https://doi.org/10.1103/PhysRevB.43.130} {\bibfield
  {journal} {\bibinfo  {journal} {Phys. Rev. B}\ }\textbf {\bibinfo {volume}
  {43}},\ \bibinfo {pages} {130} (\bibinfo {year} {1991})}\BibitemShut
  {NoStop}%
\bibitem [{\citenamefont {Blatter}\ \emph {et~al.}(1994)\citenamefont
  {Blatter}, \citenamefont {Feigel'man}, \citenamefont {Geshkenbein},
  \citenamefont {Larkin},\ and\ \citenamefont {Vinokur}}]{blatter1994vortices}%
  \BibitemOpen
  \bibfield  {author} {\bibinfo {author} {\bibfnamefont {G.}~\bibnamefont
  {Blatter}}, \bibinfo {author} {\bibfnamefont {M.~V.}\ \bibnamefont
  {Feigel'man}}, \bibinfo {author} {\bibfnamefont {V.~B.}\ \bibnamefont
  {Geshkenbein}}, \bibinfo {author} {\bibfnamefont {A.~I.}\ \bibnamefont
  {Larkin}},\ and\ \bibinfo {author} {\bibfnamefont {V.~M.}\ \bibnamefont
  {Vinokur}},\ }\href {https://doi.org/10.1103/RevModPhys.66.1125} {\bibfield
  {journal} {\bibinfo  {journal} {Rev. Mod. Phys.}\ }\textbf {\bibinfo {volume}
  {66}},\ \bibinfo {pages} {1125} (\bibinfo {year} {1994})}\BibitemShut
  {NoStop}%
\bibitem [{\citenamefont {Shibata}\ \emph {et~al.}(2002)\citenamefont
  {Shibata}, \citenamefont {Nishizaki}, \citenamefont {Sasaki},\ and\
  \citenamefont {Kobayashi}}]{shibata2002phase}%
  \BibitemOpen
  \bibfield  {author} {\bibinfo {author} {\bibfnamefont {K.}~\bibnamefont
  {Shibata}}, \bibinfo {author} {\bibfnamefont {T.}~\bibnamefont {Nishizaki}},
  \bibinfo {author} {\bibfnamefont {T.}~\bibnamefont {Sasaki}},\ and\ \bibinfo
  {author} {\bibfnamefont {N.}~\bibnamefont {Kobayashi}},\ }\href
  {https://doi.org/10.1103/PhysRevB.66.214518} {\bibfield  {journal} {\bibinfo
  {journal} {Phys. Rev. B}\ }\textbf {\bibinfo {volume} {66}},\ \bibinfo
  {pages} {214518} (\bibinfo {year} {2002})}\BibitemShut {NoStop}%
\bibitem [{\citenamefont {Salomaa}\ and\ \citenamefont
  {Volovik}(1987)}]{salomaa1987quantized}%
  \BibitemOpen
  \bibfield  {author} {\bibinfo {author} {\bibfnamefont {M.~M.}\ \bibnamefont
  {Salomaa}}\ and\ \bibinfo {author} {\bibfnamefont {G.~E.}\ \bibnamefont
  {Volovik}},\ }\href {https://doi.org/10.1103/RevModPhys.59.533} {\bibfield
  {journal} {\bibinfo  {journal} {Rev. Mod. Phys.}\ }\textbf {\bibinfo {volume}
  {59}},\ \bibinfo {pages} {533} (\bibinfo {year} {1987})}\BibitemShut
  {NoStop}%
\bibitem [{\citenamefont {Volovik}(1999)}]{volovik1999monopole}%
  \BibitemOpen
  \bibfield  {author} {\bibinfo {author} {\bibfnamefont {G.~E.}\ \bibnamefont
  {Volovik}},\ }\href {https://doi.org/10.1134/1.568231} {\bibfield  {journal}
  {\bibinfo  {journal} {J. Exp. Theor. Phys. Lett.}\ }\textbf {\bibinfo
  {volume} {70}},\ \bibinfo {pages} {792} (\bibinfo {year} {1999})}\BibitemShut
  {NoStop}%
\bibitem [{\citenamefont {Bending}(1999)}]{bending1999}%
  \BibitemOpen
  \bibfield  {author} {\bibinfo {author} {\bibfnamefont {S.~J.}\ \bibnamefont
  {Bending}},\ }\href {https://doi.org/10.1080/000187399243437} {\bibfield
  {journal} {\bibinfo  {journal} {Adv. Phys.}\ }\textbf {\bibinfo {volume}
  {48}},\ \bibinfo {pages} {449} (\bibinfo {year} {1999})}\BibitemShut
  {NoStop}%
\bibitem [{\citenamefont {Celotta}\ \emph {et~al.}(2012)\citenamefont
  {Celotta}, \citenamefont {Unguris}, \citenamefont {Kelley},\ and\
  \citenamefont {Pierce}}]{celotta2012}%
  \BibitemOpen
  \bibfield  {author} {\bibinfo {author} {\bibfnamefont {R.~J.}\ \bibnamefont
  {Celotta}}, \bibinfo {author} {\bibfnamefont {J.}~\bibnamefont {Unguris}},
  \bibinfo {author} {\bibfnamefont {M.~H.}\ \bibnamefont {Kelley}},\ and\
  \bibinfo {author} {\bibfnamefont {D.~T.}\ \bibnamefont {Pierce}},\ }\bibinfo
  {title} {Techniques to measure magnetic domain structures},\ in\ \href
  {https://doi.org/https://doi.org/10.1002/0471266965.com046.pub2} {\emph
  {\bibinfo {booktitle} {Characterization of Materials}}}\ (\bibinfo
  {publisher} {John Wiley \& Sons, Ltd},\ \bibinfo {year} {2012})\ pp.\
  \bibinfo {pages} {1--15}\BibitemShut {NoStop}%
\bibitem [{\citenamefont {Marchiori}\ \emph {et~al.}(2022)\citenamefont
  {Marchiori}, \citenamefont {Ceccarelli}, \citenamefont {Rossi}, \citenamefont
  {Lorenzelli}, \citenamefont {Degen},\ and\ \citenamefont
  {Poggio}}]{marchiori2022nanoscale}%
  \BibitemOpen
  \bibfield  {author} {\bibinfo {author} {\bibfnamefont {E.}~\bibnamefont
  {Marchiori}}, \bibinfo {author} {\bibfnamefont {L.}~\bibnamefont
  {Ceccarelli}}, \bibinfo {author} {\bibfnamefont {N.}~\bibnamefont {Rossi}},
  \bibinfo {author} {\bibfnamefont {L.}~\bibnamefont {Lorenzelli}}, \bibinfo
  {author} {\bibfnamefont {C.~L.}\ \bibnamefont {Degen}},\ and\ \bibinfo
  {author} {\bibfnamefont {M.}~\bibnamefont {Poggio}},\ }\href
  {https://doi.org/10.1038/s42254-021-00380-9} {\bibfield  {journal} {\bibinfo
  {journal} {Nat. Rev. Phys.}\ }\textbf {\bibinfo {volume} {4}},\ \bibinfo
  {pages} {49} (\bibinfo {year} {2022})}\BibitemShut {NoStop}%
\bibitem [{\citenamefont {Scholten}\ \emph {et~al.}(2021)\citenamefont
  {Scholten}, \citenamefont {Healey}, \citenamefont {Robertson}, \citenamefont
  {Abrahams}, \citenamefont {Broadway},\ and\ \citenamefont
  {Tetienne}}]{scholten2021widefield}%
  \BibitemOpen
  \bibfield  {author} {\bibinfo {author} {\bibfnamefont {S.~C.}\ \bibnamefont
  {Scholten}}, \bibinfo {author} {\bibfnamefont {A.~J.}\ \bibnamefont
  {Healey}}, \bibinfo {author} {\bibfnamefont {I.~O.}\ \bibnamefont
  {Robertson}}, \bibinfo {author} {\bibfnamefont {G.~J.}\ \bibnamefont
  {Abrahams}}, \bibinfo {author} {\bibfnamefont {D.~A.}\ \bibnamefont
  {Broadway}},\ and\ \bibinfo {author} {\bibfnamefont {J.-P.}\ \bibnamefont
  {Tetienne}},\ }\href {https://doi.org/10.1063/5.0066733} {\bibfield
  {journal} {\bibinfo  {journal} {J. Appl. Phys.}\ }\textbf {\bibinfo {volume}
  {130}},\ \bibinfo {pages} {150902} (\bibinfo {year} {2021})}\BibitemShut
  {NoStop}%
\bibitem [{\citenamefont {Kirtley}\ and\ \citenamefont
  {Wikswo~Jr}(1999)}]{kirtley1999scanning}%
  \BibitemOpen
  \bibfield  {author} {\bibinfo {author} {\bibfnamefont {J.~R.}\ \bibnamefont
  {Kirtley}}\ and\ \bibinfo {author} {\bibfnamefont {J.~P.}\ \bibnamefont
  {Wikswo~Jr}},\ }\href {https://doi.org/10.1146/annurev.matsci.29.1.117}
  {\bibfield  {journal} {\bibinfo  {journal} {Annu. Rev. Mater. Sci.}\ }\textbf
  {\bibinfo {volume} {29}},\ \bibinfo {pages} {117} (\bibinfo {year}
  {1999})}\BibitemShut {NoStop}%
\bibitem [{\citenamefont {Finkler}\ \emph {et~al.}(2012)\citenamefont
  {Finkler}, \citenamefont {Vasyukov}, \citenamefont {Segev}, \citenamefont
  {Ne{\textquotesingle}eman}, \citenamefont {Lachman}, \citenamefont
  {Rappaport}, \citenamefont {Myasoedov}, \citenamefont {Zeldov},\ and\
  \citenamefont {Huber}}]{finkler2012scanning}%
  \BibitemOpen
  \bibfield  {author} {\bibinfo {author} {\bibfnamefont {A.}~\bibnamefont
  {Finkler}}, \bibinfo {author} {\bibfnamefont {D.}~\bibnamefont {Vasyukov}},
  \bibinfo {author} {\bibfnamefont {Y.}~\bibnamefont {Segev}}, \bibinfo
  {author} {\bibfnamefont {L.}~\bibnamefont {Ne{\textquotesingle}eman}},
  \bibinfo {author} {\bibfnamefont {E.~O.}\ \bibnamefont {Lachman}}, \bibinfo
  {author} {\bibfnamefont {M.~L.}\ \bibnamefont {Rappaport}}, \bibinfo {author}
  {\bibfnamefont {Y.}~\bibnamefont {Myasoedov}}, \bibinfo {author}
  {\bibfnamefont {E.}~\bibnamefont {Zeldov}},\ and\ \bibinfo {author}
  {\bibfnamefont {M.~E.}\ \bibnamefont {Huber}},\ }\href
  {https://doi.org/10.1063/1.4731656} {\bibfield  {journal} {\bibinfo
  {journal} {Rev. Sci. Instr.}\ }\textbf {\bibinfo {volume} {83}},\ \bibinfo
  {pages} {073702} (\bibinfo {year} {2012})}\BibitemShut {NoStop}%
\bibitem [{\citenamefont {Kirtley}(2010)}]{kirtley2010fundamental}%
  \BibitemOpen
  \bibfield  {author} {\bibinfo {author} {\bibfnamefont {J.~R.}\ \bibnamefont
  {Kirtley}},\ }\href {https://dx.doi.org/10.1088/0034-4885/73/12/126501}
  {\bibfield  {journal} {\bibinfo  {journal} {Rep. Prog. Phys.}\ }\textbf
  {\bibinfo {volume} {73}},\ \bibinfo {pages} {126501} (\bibinfo {year}
  {2010})}\BibitemShut {NoStop}%
\bibitem [{\citenamefont {Thiel}\ \emph {et~al.}(2016)\citenamefont {Thiel},
  \citenamefont {Rohner}, \citenamefont {Ganzhorn}, \citenamefont {Appel},
  \citenamefont {Neu}, \citenamefont {M{\"u}ller}, \citenamefont {Kleiner},
  \citenamefont {Koelle},\ and\ \citenamefont
  {Maletinsky}}]{thiel2016quantitative}%
  \BibitemOpen
  \bibfield  {author} {\bibinfo {author} {\bibfnamefont {L.}~\bibnamefont
  {Thiel}}, \bibinfo {author} {\bibfnamefont {D.}~\bibnamefont {Rohner}},
  \bibinfo {author} {\bibfnamefont {M.}~\bibnamefont {Ganzhorn}}, \bibinfo
  {author} {\bibfnamefont {P.}~\bibnamefont {Appel}}, \bibinfo {author}
  {\bibfnamefont {E.}~\bibnamefont {Neu}}, \bibinfo {author} {\bibfnamefont
  {B.}~\bibnamefont {M{\"u}ller}}, \bibinfo {author} {\bibfnamefont
  {R.}~\bibnamefont {Kleiner}}, \bibinfo {author} {\bibfnamefont
  {D.}~\bibnamefont {Koelle}},\ and\ \bibinfo {author} {\bibfnamefont
  {P.}~\bibnamefont {Maletinsky}},\ }\href
  {https://doi.org/10.1038/nnano.2016.63} {\bibfield  {journal} {\bibinfo
  {journal} {Nat. Nanotechnol.}\ }\textbf {\bibinfo {volume} {11}},\ \bibinfo
  {pages} {677} (\bibinfo {year} {2016})}\BibitemShut {NoStop}%
\bibitem [{\citenamefont {Pelliccione}\ \emph {et~al.}(2016)\citenamefont
  {Pelliccione}, \citenamefont {Jenkins}, \citenamefont {Ovartchaiyapong},
  \citenamefont {Reetz}, \citenamefont {Emmanouilidou}, \citenamefont {Ni},\
  and\ \citenamefont {Bleszynski~Jayich}}]{pelliccione2016scanned}%
  \BibitemOpen
  \bibfield  {author} {\bibinfo {author} {\bibfnamefont {M.}~\bibnamefont
  {Pelliccione}}, \bibinfo {author} {\bibfnamefont {A.}~\bibnamefont
  {Jenkins}}, \bibinfo {author} {\bibfnamefont {P.}~\bibnamefont
  {Ovartchaiyapong}}, \bibinfo {author} {\bibfnamefont {C.}~\bibnamefont
  {Reetz}}, \bibinfo {author} {\bibfnamefont {E.}~\bibnamefont
  {Emmanouilidou}}, \bibinfo {author} {\bibfnamefont {N.}~\bibnamefont {Ni}},\
  and\ \bibinfo {author} {\bibfnamefont {A.~C.}\ \bibnamefont
  {Bleszynski~Jayich}},\ }\href {https://doi.org/10.1038/nnano.2016.68}
  {\bibfield  {journal} {\bibinfo  {journal} {Nat. Nanotechnol.}\ }\textbf
  {\bibinfo {volume} {11}},\ \bibinfo {pages} {700} (\bibinfo {year}
  {2016})}\BibitemShut {NoStop}%
\bibitem [{\citenamefont {Schirhagl}\ \emph {et~al.}(2014)\citenamefont
  {Schirhagl}, \citenamefont {Chang}, \citenamefont {Loretz},\ and\
  \citenamefont {Degen}}]{schirhagl2014nitrogen}%
  \BibitemOpen
  \bibfield  {author} {\bibinfo {author} {\bibfnamefont {R.}~\bibnamefont
  {Schirhagl}}, \bibinfo {author} {\bibfnamefont {K.}~\bibnamefont {Chang}},
  \bibinfo {author} {\bibfnamefont {M.}~\bibnamefont {Loretz}},\ and\ \bibinfo
  {author} {\bibfnamefont {C.~L.}\ \bibnamefont {Degen}},\ }\href
  {https://doi.org/10.1146/annurev-physchem-040513-103659} {\bibfield
  {journal} {\bibinfo  {journal} {Annu. Rev. Phys. Chem.}\ }\textbf {\bibinfo
  {volume} {65}},\ \bibinfo {pages} {83} (\bibinfo {year} {2014})}\BibitemShut
  {NoStop}%
\bibitem [{\citenamefont {Fu}\ \emph {et~al.}(2020)\citenamefont {Fu},
  \citenamefont {Iwata}, \citenamefont {Wickenbrock},\ and\ \citenamefont
  {Budker}}]{fu2020sensitive}%
  \BibitemOpen
  \bibfield  {author} {\bibinfo {author} {\bibfnamefont {K.-M.~C.}\
  \bibnamefont {Fu}}, \bibinfo {author} {\bibfnamefont {G.~Z.}\ \bibnamefont
  {Iwata}}, \bibinfo {author} {\bibfnamefont {A.}~\bibnamefont {Wickenbrock}},\
  and\ \bibinfo {author} {\bibfnamefont {D.}~\bibnamefont {Budker}},\ }\href
  {https://doi.org/10.1116/5.0025186} {\bibfield  {journal} {\bibinfo
  {journal} {AVS Quantum Science}\ }\textbf {\bibinfo {volume} {2}},\ \bibinfo
  {pages} {044702} (\bibinfo {year} {2020})}\BibitemShut {NoStop}%
\bibitem [{\citenamefont {Levine}\ \emph {et~al.}(2019)\citenamefont {Levine},
  \citenamefont {Turner}, \citenamefont {Kehayias}, \citenamefont {Hart},
  \citenamefont {Langellier}, \citenamefont {Trubko}, \citenamefont {Glenn},
  \citenamefont {Fu},\ and\ \citenamefont {Walsworth}}]{levine2019principles}%
  \BibitemOpen
  \bibfield  {author} {\bibinfo {author} {\bibfnamefont {E.~V.}\ \bibnamefont
  {Levine}}, \bibinfo {author} {\bibfnamefont {M.~J.}\ \bibnamefont {Turner}},
  \bibinfo {author} {\bibfnamefont {P.}~\bibnamefont {Kehayias}}, \bibinfo
  {author} {\bibfnamefont {C.~A.}\ \bibnamefont {Hart}}, \bibinfo {author}
  {\bibfnamefont {N.}~\bibnamefont {Langellier}}, \bibinfo {author}
  {\bibfnamefont {R.}~\bibnamefont {Trubko}}, \bibinfo {author} {\bibfnamefont
  {D.~R.}\ \bibnamefont {Glenn}}, \bibinfo {author} {\bibfnamefont {R.~R.}\
  \bibnamefont {Fu}},\ and\ \bibinfo {author} {\bibfnamefont {R.~L.}\
  \bibnamefont {Walsworth}},\ }\href {https://doi.org/10.1515/nanoph-2019-0209}
  {\bibfield  {journal} {\bibinfo  {journal} {Nanophotonics}\ }\textbf
  {\bibinfo {volume} {8}},\ \bibinfo {pages} {1945} (\bibinfo {year}
  {2019})}\BibitemShut {NoStop}%
\bibitem [{\citenamefont {Hsieh}\ \emph {et~al.}(2019)\citenamefont {Hsieh},
  \citenamefont {Bhattacharyya}, \citenamefont {Zu}, \citenamefont {Mittiga},
  \citenamefont {Smart}, \citenamefont {Machado}, \citenamefont {Kobrin},
  \citenamefont {H\"{o}hn}, \citenamefont {Rui}, \citenamefont {Kamrani},
  \citenamefont {Chatterjee}, \citenamefont {Choi}, \citenamefont {Zaletel},
  \citenamefont {Struzhkin}, \citenamefont {Moore}, \citenamefont {Levitas},
  \citenamefont {Jeanloz},\ and\ \citenamefont {Yao}}]{Hsieh2019}%
  \BibitemOpen
  \bibfield  {author} {\bibinfo {author} {\bibfnamefont {S.}~\bibnamefont
  {Hsieh}}, \bibinfo {author} {\bibfnamefont {P.}~\bibnamefont
  {Bhattacharyya}}, \bibinfo {author} {\bibfnamefont {C.}~\bibnamefont {Zu}},
  \bibinfo {author} {\bibfnamefont {T.}~\bibnamefont {Mittiga}}, \bibinfo
  {author} {\bibfnamefont {T.~J.}\ \bibnamefont {Smart}}, \bibinfo {author}
  {\bibfnamefont {F.}~\bibnamefont {Machado}}, \bibinfo {author} {\bibfnamefont
  {B.}~\bibnamefont {Kobrin}}, \bibinfo {author} {\bibfnamefont {T.~O.}\
  \bibnamefont {H\"{o}hn}}, \bibinfo {author} {\bibfnamefont {N.~Z.}\
  \bibnamefont {Rui}}, \bibinfo {author} {\bibfnamefont {M.}~\bibnamefont
  {Kamrani}}, \bibinfo {author} {\bibfnamefont {S.}~\bibnamefont {Chatterjee}},
  \bibinfo {author} {\bibfnamefont {S.}~\bibnamefont {Choi}}, \bibinfo {author}
  {\bibfnamefont {M.}~\bibnamefont {Zaletel}}, \bibinfo {author} {\bibfnamefont
  {V.~V.}\ \bibnamefont {Struzhkin}}, \bibinfo {author} {\bibfnamefont {J.~E.}\
  \bibnamefont {Moore}}, \bibinfo {author} {\bibfnamefont {V.~I.}\ \bibnamefont
  {Levitas}}, \bibinfo {author} {\bibfnamefont {R.}~\bibnamefont {Jeanloz}},\
  and\ \bibinfo {author} {\bibfnamefont {N.~Y.}\ \bibnamefont {Yao}},\ }\href
  {https://doi.org/10.1126/science.aaw4352} {\bibfield  {journal} {\bibinfo
  {journal} {Science}\ }\textbf {\bibinfo {volume} {366}},\ \bibinfo {pages}
  {1349} (\bibinfo {year} {2019})}\BibitemShut {NoStop}%
\bibitem [{\citenamefont {Lesik}\ \emph {et~al.}(2019)\citenamefont {Lesik},
  \citenamefont {Plisson}, \citenamefont {Toraille}, \citenamefont {Renaud},
  \citenamefont {Occelli}, \citenamefont {Schmidt}, \citenamefont {Salord},
  \citenamefont {Delobbe}, \citenamefont {Debuisschert}, \citenamefont
  {Rondin}, \citenamefont {Loubeyre},\ and\ \citenamefont {Roch}}]{Lesik2019}%
  \BibitemOpen
  \bibfield  {author} {\bibinfo {author} {\bibfnamefont {M.}~\bibnamefont
  {Lesik}}, \bibinfo {author} {\bibfnamefont {T.}~\bibnamefont {Plisson}},
  \bibinfo {author} {\bibfnamefont {L.}~\bibnamefont {Toraille}}, \bibinfo
  {author} {\bibfnamefont {J.}~\bibnamefont {Renaud}}, \bibinfo {author}
  {\bibfnamefont {F.}~\bibnamefont {Occelli}}, \bibinfo {author} {\bibfnamefont
  {M.}~\bibnamefont {Schmidt}}, \bibinfo {author} {\bibfnamefont
  {O.}~\bibnamefont {Salord}}, \bibinfo {author} {\bibfnamefont
  {A.}~\bibnamefont {Delobbe}}, \bibinfo {author} {\bibfnamefont
  {T.}~\bibnamefont {Debuisschert}}, \bibinfo {author} {\bibfnamefont
  {L.}~\bibnamefont {Rondin}}, \bibinfo {author} {\bibfnamefont
  {P.}~\bibnamefont {Loubeyre}},\ and\ \bibinfo {author} {\bibfnamefont
  {J.-F.}\ \bibnamefont {Roch}},\ }\href
  {https://doi.org/10.1126/science.aaw4329} {\bibfield  {journal} {\bibinfo
  {journal} {Science}\ }\textbf {\bibinfo {volume} {366}},\ \bibinfo {pages}
  {1359} (\bibinfo {year} {2019})}\BibitemShut {NoStop}%
\bibitem [{\citenamefont {Toraille}\ \emph {et~al.}(2020)\citenamefont
  {Toraille}, \citenamefont {Hilberer}, \citenamefont {Plisson}, \citenamefont
  {Lesik}, \citenamefont {Chipaux}, \citenamefont {Vindolet}, \citenamefont
  {P{\'{e}}pin}, \citenamefont {Occelli}, \citenamefont {Schmidt},
  \citenamefont {Debuisschert}, \citenamefont {Guignot}, \citenamefont
  {Iti{\'{e}}}, \citenamefont {Loubeyre},\ and\ \citenamefont
  {Roch}}]{Toraille2020}%
  \BibitemOpen
  \bibfield  {author} {\bibinfo {author} {\bibfnamefont {L.}~\bibnamefont
  {Toraille}}, \bibinfo {author} {\bibfnamefont {A.}~\bibnamefont {Hilberer}},
  \bibinfo {author} {\bibfnamefont {T.}~\bibnamefont {Plisson}}, \bibinfo
  {author} {\bibfnamefont {M.}~\bibnamefont {Lesik}}, \bibinfo {author}
  {\bibfnamefont {M.}~\bibnamefont {Chipaux}}, \bibinfo {author} {\bibfnamefont
  {B.}~\bibnamefont {Vindolet}}, \bibinfo {author} {\bibfnamefont
  {C.}~\bibnamefont {P{\'{e}}pin}}, \bibinfo {author} {\bibfnamefont
  {F.}~\bibnamefont {Occelli}}, \bibinfo {author} {\bibfnamefont
  {M.}~\bibnamefont {Schmidt}}, \bibinfo {author} {\bibfnamefont
  {T.}~\bibnamefont {Debuisschert}}, \bibinfo {author} {\bibfnamefont
  {N.}~\bibnamefont {Guignot}}, \bibinfo {author} {\bibfnamefont {J.-P.}\
  \bibnamefont {Iti{\'{e}}}}, \bibinfo {author} {\bibfnamefont
  {P.}~\bibnamefont {Loubeyre}},\ and\ \bibinfo {author} {\bibfnamefont
  {J.-F.}\ \bibnamefont {Roch}},\ }\href
  {https://doi.org/10.1088/1367-2630/abc28f} {\bibfield  {journal} {\bibinfo
  {journal} {New J. Phys.}\ }\textbf {\bibinfo {volume} {22}},\ \bibinfo
  {pages} {103063} (\bibinfo {year} {2020})}\BibitemShut {NoStop}%
\bibitem [{\citenamefont {Drozdov}\ \emph {et~al.}(2015)\citenamefont
  {Drozdov}, \citenamefont {Eremets}, \citenamefont {Troyan}, \citenamefont
  {Ksenofontov},\ and\ \citenamefont {Shylin}}]{drozdov2015conventional}%
  \BibitemOpen
  \bibfield  {author} {\bibinfo {author} {\bibfnamefont {A.~P.}\ \bibnamefont
  {Drozdov}}, \bibinfo {author} {\bibfnamefont {M.~I.}\ \bibnamefont
  {Eremets}}, \bibinfo {author} {\bibfnamefont {I.~A.}\ \bibnamefont {Troyan}},
  \bibinfo {author} {\bibfnamefont {V.}~\bibnamefont {Ksenofontov}},\ and\
  \bibinfo {author} {\bibfnamefont {S.~I.}\ \bibnamefont {Shylin}},\ }\href
  {https://doi.org/10.1038/nature14964} {\bibfield  {journal} {\bibinfo
  {journal} {Nature}\ }\textbf {\bibinfo {volume} {525}},\ \bibinfo {pages}
  {73} (\bibinfo {year} {2015})}\BibitemShut {NoStop}%
\bibitem [{\citenamefont {Mozaffari}\ \emph {et~al.}(2019)\citenamefont
  {Mozaffari}, \citenamefont {Sun}, \citenamefont {Minkov}, \citenamefont
  {Drozdov}, \citenamefont {Knyazev}, \citenamefont {Betts}, \citenamefont
  {Einaga}, \citenamefont {Shimizu}, \citenamefont {Eremets}, \citenamefont
  {Balicas},\ and\ \citenamefont {Balakirev}}]{mozaffari2019superconducting}%
  \BibitemOpen
  \bibfield  {author} {\bibinfo {author} {\bibfnamefont {S.}~\bibnamefont
  {Mozaffari}}, \bibinfo {author} {\bibfnamefont {D.}~\bibnamefont {Sun}},
  \bibinfo {author} {\bibfnamefont {V.~S.}\ \bibnamefont {Minkov}}, \bibinfo
  {author} {\bibfnamefont {A.~P.}\ \bibnamefont {Drozdov}}, \bibinfo {author}
  {\bibfnamefont {D.}~\bibnamefont {Knyazev}}, \bibinfo {author} {\bibfnamefont
  {J.~B.}\ \bibnamefont {Betts}}, \bibinfo {author} {\bibfnamefont
  {M.}~\bibnamefont {Einaga}}, \bibinfo {author} {\bibfnamefont
  {K.}~\bibnamefont {Shimizu}}, \bibinfo {author} {\bibfnamefont {M.~I.}\
  \bibnamefont {Eremets}}, \bibinfo {author} {\bibfnamefont {L.}~\bibnamefont
  {Balicas}},\ and\ \bibinfo {author} {\bibfnamefont {F.~F.}\ \bibnamefont
  {Balakirev}},\ }\href {https://doi.org/10.1038/s41467-019-10552-y} {\bibfield
   {journal} {\bibinfo  {journal} {Nat. Commn.}\ }\textbf {\bibinfo {volume}
  {10}},\ \bibinfo {pages} {2522} (\bibinfo {year} {2019})}\BibitemShut
  {NoStop}%
\bibitem [{\citenamefont {Schlussel}\ \emph {et~al.}(2018)\citenamefont
  {Schlussel}, \citenamefont {Lenz}, \citenamefont {Rohner}, \citenamefont
  {Bar-Haim}, \citenamefont {Bougas}, \citenamefont {Groswasser}, \citenamefont
  {Kieschnick}, \citenamefont {Rozenberg}, \citenamefont {Thiel}, \citenamefont
  {Waxman}, \citenamefont {Meijer}, \citenamefont {Maletinsky}, \citenamefont
  {Budker},\ and\ \citenamefont {Folman}}]{schlussel2018wide}%
  \BibitemOpen
  \bibfield  {author} {\bibinfo {author} {\bibfnamefont {Y.}~\bibnamefont
  {Schlussel}}, \bibinfo {author} {\bibfnamefont {T.}~\bibnamefont {Lenz}},
  \bibinfo {author} {\bibfnamefont {D.}~\bibnamefont {Rohner}}, \bibinfo
  {author} {\bibfnamefont {Y.}~\bibnamefont {Bar-Haim}}, \bibinfo {author}
  {\bibfnamefont {L.}~\bibnamefont {Bougas}}, \bibinfo {author} {\bibfnamefont
  {D.}~\bibnamefont {Groswasser}}, \bibinfo {author} {\bibfnamefont
  {M.}~\bibnamefont {Kieschnick}}, \bibinfo {author} {\bibfnamefont
  {E.}~\bibnamefont {Rozenberg}}, \bibinfo {author} {\bibfnamefont
  {L.}~\bibnamefont {Thiel}}, \bibinfo {author} {\bibfnamefont
  {A.}~\bibnamefont {Waxman}}, \bibinfo {author} {\bibfnamefont
  {J.}~\bibnamefont {Meijer}}, \bibinfo {author} {\bibfnamefont
  {P.}~\bibnamefont {Maletinsky}}, \bibinfo {author} {\bibfnamefont
  {D.}~\bibnamefont {Budker}},\ and\ \bibinfo {author} {\bibfnamefont
  {R.}~\bibnamefont {Folman}},\ }\href
  {https://doi.org/10.1103/physrevapplied.10.034032} {\bibfield  {journal}
  {\bibinfo  {journal} {Phys. Rev. Appl.}\ }\textbf {\bibinfo {volume} {10}},\
  \bibinfo {pages} {034032} (\bibinfo {year} {2018})}\BibitemShut {NoStop}%
\bibitem [{\citenamefont {Lillie}\ \emph {et~al.}(2020)\citenamefont {Lillie},
  \citenamefont {Broadway}, \citenamefont {Dontschuk}, \citenamefont
  {Scholten}, \citenamefont {Johnson}, \citenamefont {Wolf}, \citenamefont
  {Rachel}, \citenamefont {Hollenberg},\ and\ \citenamefont
  {Tetienne}}]{scott2020}%
  \BibitemOpen
  \bibfield  {author} {\bibinfo {author} {\bibfnamefont {S.~E.}\ \bibnamefont
  {Lillie}}, \bibinfo {author} {\bibfnamefont {D.~A.}\ \bibnamefont
  {Broadway}}, \bibinfo {author} {\bibfnamefont {N.}~\bibnamefont {Dontschuk}},
  \bibinfo {author} {\bibfnamefont {S.~C.}\ \bibnamefont {Scholten}}, \bibinfo
  {author} {\bibfnamefont {B.~C.}\ \bibnamefont {Johnson}}, \bibinfo {author}
  {\bibfnamefont {S.}~\bibnamefont {Wolf}}, \bibinfo {author} {\bibfnamefont
  {S.}~\bibnamefont {Rachel}}, \bibinfo {author} {\bibfnamefont {L.~C.~L.}\
  \bibnamefont {Hollenberg}},\ and\ \bibinfo {author} {\bibfnamefont {J.-P.}\
  \bibnamefont {Tetienne}},\ }\href
  {https://doi.org/10.1021/acs.nanolett.9b05071} {\bibfield  {journal}
  {\bibinfo  {journal} {Nano Letters}\ }\textbf {\bibinfo {volume} {20}},\
  \bibinfo {pages} {1855} (\bibinfo {year} {2020})}\BibitemShut {NoStop}%
\bibitem [{\citenamefont {Rondin}\ \emph {et~al.}(2014)\citenamefont {Rondin},
  \citenamefont {Tetienne}, \citenamefont {Hingant}, \citenamefont {Roch},
  \citenamefont {Maletinsky},\ and\ \citenamefont
  {Jacques}}]{rondin2014magnetometry}%
  \BibitemOpen
  \bibfield  {author} {\bibinfo {author} {\bibfnamefont {L.}~\bibnamefont
  {Rondin}}, \bibinfo {author} {\bibfnamefont {J.-P.}\ \bibnamefont
  {Tetienne}}, \bibinfo {author} {\bibfnamefont {T.}~\bibnamefont {Hingant}},
  \bibinfo {author} {\bibfnamefont {J.-F.}\ \bibnamefont {Roch}}, \bibinfo
  {author} {\bibfnamefont {P.}~\bibnamefont {Maletinsky}},\ and\ \bibinfo
  {author} {\bibfnamefont {V.}~\bibnamefont {Jacques}},\ }\href
  {https://doi.org/10.1088/0034-4885/77/5/056503} {\bibfield  {journal}
  {\bibinfo  {journal} {Rep. Prog. Phys.}\ }\textbf {\bibinfo {volume} {77}},\
  \bibinfo {pages} {056503} (\bibinfo {year} {2014})}\BibitemShut {NoStop}%
\bibitem [{\citenamefont {Miyazaki}\ \emph {et~al.}(2014)\citenamefont
  {Miyazaki}, \citenamefont {Miyamoto}, \citenamefont {Makino}, \citenamefont
  {Kato}, \citenamefont {Yamasaki}, \citenamefont {Fukui}, \citenamefont {Doi},
  \citenamefont {Tokuda}, \citenamefont {Hatano},\ and\ \citenamefont
  {Mizuochi}}]{miyazaki2014atomistic}%
  \BibitemOpen
  \bibfield  {author} {\bibinfo {author} {\bibfnamefont {T.}~\bibnamefont
  {Miyazaki}}, \bibinfo {author} {\bibfnamefont {Y.}~\bibnamefont {Miyamoto}},
  \bibinfo {author} {\bibfnamefont {T.}~\bibnamefont {Makino}}, \bibinfo
  {author} {\bibfnamefont {H.}~\bibnamefont {Kato}}, \bibinfo {author}
  {\bibfnamefont {S.}~\bibnamefont {Yamasaki}}, \bibinfo {author}
  {\bibfnamefont {T.}~\bibnamefont {Fukui}}, \bibinfo {author} {\bibfnamefont
  {Y.}~\bibnamefont {Doi}}, \bibinfo {author} {\bibfnamefont {N.}~\bibnamefont
  {Tokuda}}, \bibinfo {author} {\bibfnamefont {M.}~\bibnamefont {Hatano}},\
  and\ \bibinfo {author} {\bibfnamefont {N.}~\bibnamefont {Mizuochi}},\ }\href
  {https://doi.org/10.1063/1.4904988} {\bibfield  {journal} {\bibinfo
  {journal} {Appl. Phys. Lett.}\ }\textbf {\bibinfo {volume} {105}},\ \bibinfo
  {pages} {261601} (\bibinfo {year} {2014})}\BibitemShut {NoStop}%
\bibitem [{\citenamefont {Tahara}\ \emph {et~al.}(2015)\citenamefont {Tahara},
  \citenamefont {Ozawa}, \citenamefont {Iwasaki}, \citenamefont {Mizuochi},\
  and\ \citenamefont {Hatano}}]{tahara2015quantifying}%
  \BibitemOpen
  \bibfield  {author} {\bibinfo {author} {\bibfnamefont {K.}~\bibnamefont
  {Tahara}}, \bibinfo {author} {\bibfnamefont {H.}~\bibnamefont {Ozawa}},
  \bibinfo {author} {\bibfnamefont {T.}~\bibnamefont {Iwasaki}}, \bibinfo
  {author} {\bibfnamefont {N.}~\bibnamefont {Mizuochi}},\ and\ \bibinfo
  {author} {\bibfnamefont {M.}~\bibnamefont {Hatano}},\ }\href
  {https://doi.org/10.1063/1.4935709} {\bibfield  {journal} {\bibinfo
  {journal} {Appl. Phys. Lett.}\ }\textbf {\bibinfo {volume} {107}},\ \bibinfo
  {pages} {193110} (\bibinfo {year} {2015})}\BibitemShut {NoStop}%
\bibitem [{\citenamefont {Ishiwata}\ \emph {et~al.}(2017)\citenamefont
  {Ishiwata}, \citenamefont {Nakajima}, \citenamefont {Tahara}, \citenamefont
  {Ozawa}, \citenamefont {Iwasaki},\ and\ \citenamefont
  {Hatano}}]{ishiwata2017perfectly}%
  \BibitemOpen
  \bibfield  {author} {\bibinfo {author} {\bibfnamefont {H.}~\bibnamefont
  {Ishiwata}}, \bibinfo {author} {\bibfnamefont {M.}~\bibnamefont {Nakajima}},
  \bibinfo {author} {\bibfnamefont {K.}~\bibnamefont {Tahara}}, \bibinfo
  {author} {\bibfnamefont {H.}~\bibnamefont {Ozawa}}, \bibinfo {author}
  {\bibfnamefont {T.}~\bibnamefont {Iwasaki}},\ and\ \bibinfo {author}
  {\bibfnamefont {M.}~\bibnamefont {Hatano}},\ }\href
  {https://doi.org/10.1063/1.4993160} {\bibfield  {journal} {\bibinfo
  {journal} {Appl. Phys. Lett.}\ }\textbf {\bibinfo {volume} {111}},\ \bibinfo
  {pages} {043103} (\bibinfo {year} {2017})}\BibitemShut {NoStop}%
\bibitem [{\citenamefont {Ozawa}\ \emph {et~al.}(2017)\citenamefont {Ozawa},
  \citenamefont {Tahara}, \citenamefont {Ishiwata}, \citenamefont {Hatano},\
  and\ \citenamefont {Iwasaki}}]{ozawa2017formation}%
  \BibitemOpen
  \bibfield  {author} {\bibinfo {author} {\bibfnamefont {H.}~\bibnamefont
  {Ozawa}}, \bibinfo {author} {\bibfnamefont {K.}~\bibnamefont {Tahara}},
  \bibinfo {author} {\bibfnamefont {H.}~\bibnamefont {Ishiwata}}, \bibinfo
  {author} {\bibfnamefont {M.}~\bibnamefont {Hatano}},\ and\ \bibinfo {author}
  {\bibfnamefont {T.}~\bibnamefont {Iwasaki}},\ }\href
  {https://doi.org/10.7567/APEX.10.045501} {\bibfield  {journal} {\bibinfo
  {journal} {Appl. Phys. Express}\ }\textbf {\bibinfo {volume} {10}},\ \bibinfo
  {pages} {045501} (\bibinfo {year} {2017})}\BibitemShut {NoStop}%
\bibitem [{\citenamefont {Tsuji}\ \emph {et~al.}(2022)\citenamefont {Tsuji},
  \citenamefont {Ishiwata}, \citenamefont {Sekiguchi}, \citenamefont
  {Iwasaki},\ and\ \citenamefont {Hatano}}]{Tsuji2022}%
  \BibitemOpen
  \bibfield  {author} {\bibinfo {author} {\bibfnamefont {T.}~\bibnamefont
  {Tsuji}}, \bibinfo {author} {\bibfnamefont {H.}~\bibnamefont {Ishiwata}},
  \bibinfo {author} {\bibfnamefont {T.}~\bibnamefont {Sekiguchi}}, \bibinfo
  {author} {\bibfnamefont {T.}~\bibnamefont {Iwasaki}},\ and\ \bibinfo {author}
  {\bibfnamefont {M.}~\bibnamefont {Hatano}},\ }\href
  {https://doi.org/10.1016/j.diamond.2022.108840} {\bibfield  {journal}
  {\bibinfo  {journal} {Diam. Relat. Mater.}\ }\textbf {\bibinfo {volume}
  {123}},\ \bibinfo {pages} {108840} (\bibinfo {year} {2022})}\BibitemShut
  {NoStop}%
\bibitem [{\citenamefont {Stewart}(2017)}]{stewart2017unconventional}%
  \BibitemOpen
  \bibfield  {author} {\bibinfo {author} {\bibfnamefont {G.~R.}\ \bibnamefont
  {Stewart}},\ }\href {https://doi.org/10.1080/00018732.2017.1331615}
  {\bibfield  {journal} {\bibinfo  {journal} {Adv. Phys.}\ }\textbf {\bibinfo
  {volume} {66}},\ \bibinfo {pages} {75} (\bibinfo {year} {2017})}\BibitemShut
  {NoStop}%
\bibitem [{\citenamefont {Sarma}\ \emph {et~al.}(2006)\citenamefont {Sarma},
  \citenamefont {Nayak},\ and\ \citenamefont {Tewari}}]{sarma2006proposal}%
  \BibitemOpen
  \bibfield  {author} {\bibinfo {author} {\bibfnamefont {S.~D.}\ \bibnamefont
  {Sarma}}, \bibinfo {author} {\bibfnamefont {C.}~\bibnamefont {Nayak}},\ and\
  \bibinfo {author} {\bibfnamefont {S.}~\bibnamefont {Tewari}},\ }\href
  {https://link.aps.org/doi/10.1103/PhysRevB.73.220502} {\bibfield  {journal}
  {\bibinfo  {journal} {Phys. Rev. B}\ }\textbf {\bibinfo {volume} {73}},\
  \bibinfo {pages} {220502} (\bibinfo {year} {2006})}\BibitemShut {NoStop}%
\bibitem [{\citenamefont {How}\ and\ \citenamefont {Yip}(2020)}]{how2020half}%
  \BibitemOpen
  \bibfield  {author} {\bibinfo {author} {\bibfnamefont {P.~T.}\ \bibnamefont
  {How}}\ and\ \bibinfo {author} {\bibfnamefont {S.-K.}\ \bibnamefont {Yip}},\
  }\href {https://link.aps.org/doi/10.1103/PhysRevResearch.2.043192} {\bibfield
   {journal} {\bibinfo  {journal} {Phys. Rev. Research}\ }\textbf {\bibinfo
  {volume} {2}},\ \bibinfo {pages} {043192} (\bibinfo {year}
  {2020})}\BibitemShut {NoStop}%
\bibitem [{\citenamefont {Nishimura}\ \emph {et~al.}()\citenamefont
  {Nishimura}, \citenamefont {Tsukamoto}, \citenamefont {Sasaki},\ and\
  \citenamefont {Kobayashi}}]{nishimura2023tobesubmitted}%
  \BibitemOpen
  \bibfield  {author} {\bibinfo {author} {\bibfnamefont {S.}~\bibnamefont
  {Nishimura}}, \bibinfo {author} {\bibfnamefont {M.}~\bibnamefont
  {Tsukamoto}}, \bibinfo {author} {\bibfnamefont {K.}~\bibnamefont {Sasaki}},\
  and\ \bibinfo {author} {\bibfnamefont {K.}~\bibnamefont {Kobayashi}},\
  }\href@noop {} {}\bibinfo {note} {In preparation}\BibitemShut {NoStop}%
\bibitem [{\citenamefont {Sasaki}\ \emph {et~al.}(2016)\citenamefont {Sasaki},
  \citenamefont {Monnai}, \citenamefont {Saijo}, \citenamefont {Fujita},
  \citenamefont {Watanabe}, \citenamefont {Ishi-Hayase}, \citenamefont {Itoh},\
  and\ \citenamefont {Abe}}]{Sasaki2016}%
  \BibitemOpen
  \bibfield  {author} {\bibinfo {author} {\bibfnamefont {K.}~\bibnamefont
  {Sasaki}}, \bibinfo {author} {\bibfnamefont {Y.}~\bibnamefont {Monnai}},
  \bibinfo {author} {\bibfnamefont {S.}~\bibnamefont {Saijo}}, \bibinfo
  {author} {\bibfnamefont {R.}~\bibnamefont {Fujita}}, \bibinfo {author}
  {\bibfnamefont {H.}~\bibnamefont {Watanabe}}, \bibinfo {author}
  {\bibfnamefont {J.}~\bibnamefont {Ishi-Hayase}}, \bibinfo {author}
  {\bibfnamefont {K.~M.}\ \bibnamefont {Itoh}},\ and\ \bibinfo {author}
  {\bibfnamefont {E.}~\bibnamefont {Abe}},\ }\href
  {https://doi.org/10.1063/1.4952418} {\bibfield  {journal} {\bibinfo
  {journal} {Rev. Sci. Instr.}\ }\textbf {\bibinfo {volume} {87}},\ \bibinfo
  {pages} {053904} (\bibinfo {year} {2016})}\BibitemShut {NoStop}%
\bibitem [{\citenamefont {Tsukamoto}\ \emph {et~al.}(2021)\citenamefont
  {Tsukamoto}, \citenamefont {Ogawa}, \citenamefont {Ozawa}, \citenamefont
  {Iwasaki}, \citenamefont {Hatano}, \citenamefont {Sasaki},\ and\
  \citenamefont {Kobayashi}}]{tsukamoto2021vector}%
  \BibitemOpen
  \bibfield  {author} {\bibinfo {author} {\bibfnamefont {M.}~\bibnamefont
  {Tsukamoto}}, \bibinfo {author} {\bibfnamefont {K.}~\bibnamefont {Ogawa}},
  \bibinfo {author} {\bibfnamefont {H.}~\bibnamefont {Ozawa}}, \bibinfo
  {author} {\bibfnamefont {T.}~\bibnamefont {Iwasaki}}, \bibinfo {author}
  {\bibfnamefont {M.}~\bibnamefont {Hatano}}, \bibinfo {author} {\bibfnamefont
  {K.}~\bibnamefont {Sasaki}},\ and\ \bibinfo {author} {\bibfnamefont
  {K.}~\bibnamefont {Kobayashi}},\ }\href {https://doi.org/10.1063/5.0054809}
  {\bibfield  {journal} {\bibinfo  {journal} {Appl. Phys. Lett.}\ }\textbf
  {\bibinfo {volume} {118}},\ \bibinfo {pages} {264002} (\bibinfo {year}
  {2021})}\BibitemShut {NoStop}%
\bibitem [{\citenamefont {Fujiwara}\ \emph {et~al.}(2020)\citenamefont
  {Fujiwara}, \citenamefont {Dohms}, \citenamefont {Suto}, \citenamefont
  {Nishimura}, \citenamefont {Oshimi}, \citenamefont {Teki}, \citenamefont
  {Cai}, \citenamefont {Benson},\ and\ \citenamefont {Shikano}}]{Fujiwara2020}%
  \BibitemOpen
  \bibfield  {author} {\bibinfo {author} {\bibfnamefont {M.}~\bibnamefont
  {Fujiwara}}, \bibinfo {author} {\bibfnamefont {A.}~\bibnamefont {Dohms}},
  \bibinfo {author} {\bibfnamefont {K.}~\bibnamefont {Suto}}, \bibinfo {author}
  {\bibfnamefont {Y.}~\bibnamefont {Nishimura}}, \bibinfo {author}
  {\bibfnamefont {K.}~\bibnamefont {Oshimi}}, \bibinfo {author} {\bibfnamefont
  {Y.}~\bibnamefont {Teki}}, \bibinfo {author} {\bibfnamefont {K.}~\bibnamefont
  {Cai}}, \bibinfo {author} {\bibfnamefont {O.}~\bibnamefont {Benson}},\ and\
  \bibinfo {author} {\bibfnamefont {Y.}~\bibnamefont {Shikano}},\ }\href
  {https://doi.org/10.1103/physrevresearch.2.043415} {\bibfield  {journal}
  {\bibinfo  {journal} {Phys. Rev. Research}\ }\textbf {\bibinfo {volume}
  {2}},\ \bibinfo {pages} {043415} (\bibinfo {year} {2020})}\BibitemShut
  {NoStop}%
\bibitem [{\citenamefont {Ito}\ \emph {et~al.}(2023)\citenamefont {Ito},
  \citenamefont {Tsukamoto}, \citenamefont {Ogawa}, \citenamefont {Teraji},
  \citenamefont {Sasaki},\ and\ \citenamefont {Kobayashi}}]{itoh2023}%
  \BibitemOpen
  \bibfield  {author} {\bibinfo {author} {\bibfnamefont {S.}~\bibnamefont
  {Ito}}, \bibinfo {author} {\bibfnamefont {M.}~\bibnamefont {Tsukamoto}},
  \bibinfo {author} {\bibfnamefont {K.}~\bibnamefont {Ogawa}}, \bibinfo
  {author} {\bibfnamefont {T.}~\bibnamefont {Teraji}}, \bibinfo {author}
  {\bibfnamefont {K.}~\bibnamefont {Sasaki}},\ and\ \bibinfo {author}
  {\bibfnamefont {K.}~\bibnamefont {Kobayashi}},\ }\href
  {https://doi.org/10.7566/JPSJ.92.084701} {\bibfield  {journal} {\bibinfo
  {journal} {Journal of the Physical Society of Japan}\ }\textbf {\bibinfo
  {volume} {92}},\ \bibinfo {pages} {084701} (\bibinfo {year}
  {2023})}\BibitemShut {NoStop}%
\bibitem [{\citenamefont {Clem}(1975)}]{clem1975}%
  \BibitemOpen
  \bibfield  {author} {\bibinfo {author} {\bibfnamefont {J.~R.}\ \bibnamefont
  {Clem}},\ }\href {https://doi.org/10.1007/BF00116134} {\bibfield  {journal}
  {\bibinfo  {journal} {Journal of Low Temperature Physics}\ }\textbf {\bibinfo
  {volume} {18}},\ \bibinfo {pages} {427} (\bibinfo {year} {1975})}\BibitemShut
  {NoStop}%
\bibitem [{\citenamefont {Carneiro}\ and\ \citenamefont
  {Brandt}(2000)}]{carneiro2000vortex}%
  \BibitemOpen
  \bibfield  {author} {\bibinfo {author} {\bibfnamefont {G.}~\bibnamefont
  {Carneiro}}\ and\ \bibinfo {author} {\bibfnamefont {E.~H.}\ \bibnamefont
  {Brandt}},\ }\href {https://doi.org/10.1103/PhysRevB.61.6370} {\bibfield
  {journal} {\bibinfo  {journal} {Phys. Rev. B}\ }\textbf {\bibinfo {volume}
  {61}},\ \bibinfo {pages} {6370} (\bibinfo {year} {2000})}\BibitemShut
  {NoStop}%
\bibitem [{\citenamefont {Kogan}(2003)}]{kogan2003}%
  \BibitemOpen
  \bibfield  {author} {\bibinfo {author} {\bibfnamefont {V.~G.}\ \bibnamefont
  {Kogan}},\ }\href {https://doi.org/10.1103/PhysRevB.68.104511} {\bibfield
  {journal} {\bibinfo  {journal} {Phys. Rev. B}\ }\textbf {\bibinfo {volume}
  {68}},\ \bibinfo {pages} {104511} (\bibinfo {year} {2003})}\BibitemShut
  {NoStop}%
\bibitem [{\citenamefont {Kogan}\ \emph {et~al.}(2021)\citenamefont {Kogan},
  \citenamefont {Nakagawa},\ and\ \citenamefont {Kirtley}}]{kogan2021}%
  \BibitemOpen
  \bibfield  {author} {\bibinfo {author} {\bibfnamefont {V.~G.}\ \bibnamefont
  {Kogan}}, \bibinfo {author} {\bibfnamefont {N.}~\bibnamefont {Nakagawa}},\
  and\ \bibinfo {author} {\bibfnamefont {J.~R.}\ \bibnamefont {Kirtley}},\
  }\href {https://doi.org/10.1103/PhysRevB.104.144512} {\bibfield  {journal}
  {\bibinfo  {journal} {Phys. Rev. B}\ }\textbf {\bibinfo {volume} {104}},\
  \bibinfo {pages} {144512} (\bibinfo {year} {2021})}\BibitemShut {NoStop}%
\bibitem [{\citenamefont {Pearl}(1964)}]{Pearl1964}%
  \BibitemOpen
  \bibfield  {author} {\bibinfo {author} {\bibfnamefont {J.}~\bibnamefont
  {Pearl}},\ }\href {https://doi.org/10.1063/1.1754056} {\bibfield  {journal}
  {\bibinfo  {journal} {Appl. Phys. Lett.}\ }\textbf {\bibinfo {volume} {5}},\
  \bibinfo {pages} {65} (\bibinfo {year} {1964})}\BibitemShut {NoStop}%
\bibitem [{\citenamefont {Djordjevic}\ \emph {et~al.}(2002)\citenamefont
  {Djordjevic}, \citenamefont {Farber}, \citenamefont {Deutscher},
  \citenamefont {Bontemps}, \citenamefont {Durand},\ and\ \citenamefont
  {Contour}}]{Djordjevic2002}%
  \BibitemOpen
  \bibfield  {author} {\bibinfo {author} {\bibfnamefont {S.}~\bibnamefont
  {Djordjevic}}, \bibinfo {author} {\bibfnamefont {E.}~\bibnamefont {Farber}},
  \bibinfo {author} {\bibfnamefont {G.}~\bibnamefont {Deutscher}}, \bibinfo
  {author} {\bibfnamefont {N.}~\bibnamefont {Bontemps}}, \bibinfo {author}
  {\bibfnamefont {O.}~\bibnamefont {Durand}},\ and\ \bibinfo {author}
  {\bibfnamefont {J.}~\bibnamefont {Contour}},\ }\href
  {https://doi.org/10.1140/epjb/e20020046} {\bibfield  {journal} {\bibinfo
  {journal} {Eur. Phys. J. B}\ }\textbf {\bibinfo {volume} {25}},\ \bibinfo
  {pages} {407} (\bibinfo {year} {2002})}\BibitemShut {NoStop}%
\bibitem [{\citenamefont {Sonier}\ \emph {et~al.}(2007)\citenamefont {Sonier},
  \citenamefont {Sabok-Sayr}, \citenamefont {Callaghan}, \citenamefont
  {Kaiser}, \citenamefont {Pacradouni}, \citenamefont {Brewer}, \citenamefont
  {Stubbs}, \citenamefont {Hardy}, \citenamefont {Bonn}, \citenamefont
  {Liang},\ and\ \citenamefont {Atkinson}}]{Sonier2007}%
  \BibitemOpen
  \bibfield  {author} {\bibinfo {author} {\bibfnamefont {J.~E.}\ \bibnamefont
  {Sonier}}, \bibinfo {author} {\bibfnamefont {S.~A.}\ \bibnamefont
  {Sabok-Sayr}}, \bibinfo {author} {\bibfnamefont {F.~D.}\ \bibnamefont
  {Callaghan}}, \bibinfo {author} {\bibfnamefont {C.~V.}\ \bibnamefont
  {Kaiser}}, \bibinfo {author} {\bibfnamefont {V.}~\bibnamefont {Pacradouni}},
  \bibinfo {author} {\bibfnamefont {J.~H.}\ \bibnamefont {Brewer}}, \bibinfo
  {author} {\bibfnamefont {S.~L.}\ \bibnamefont {Stubbs}}, \bibinfo {author}
  {\bibfnamefont {W.~N.}\ \bibnamefont {Hardy}}, \bibinfo {author}
  {\bibfnamefont {D.~A.}\ \bibnamefont {Bonn}}, \bibinfo {author}
  {\bibfnamefont {R.}~\bibnamefont {Liang}},\ and\ \bibinfo {author}
  {\bibfnamefont {W.~A.}\ \bibnamefont {Atkinson}},\ }\href
  {https://link.aps.org/doi/10.1103/PhysRevB.76.134518} {\bibfield  {journal}
  {\bibinfo  {journal} {Phys. Rev. B}\ }\textbf {\bibinfo {volume} {76}},\
  \bibinfo {pages} {134518} (\bibinfo {year} {2007})}\BibitemShut {NoStop}%
\bibitem [{\citenamefont {Hassan}\ \emph {et~al.}(2021)\citenamefont {Hassan},
  \citenamefont {Labrag}, \citenamefont {Taoufik}, \citenamefont {Bghour},
  \citenamefont {Ouaddi}, \citenamefont {Tirbiyine}, \citenamefont {Lmouden},
  \citenamefont {Hafid},\ and\ \citenamefont {Hamidi}}]{abou2021magnetic}%
  \BibitemOpen
  \bibfield  {author} {\bibinfo {author} {\bibfnamefont {A.~A.~E.}\
  \bibnamefont {Hassan}}, \bibinfo {author} {\bibfnamefont {A.}~\bibnamefont
  {Labrag}}, \bibinfo {author} {\bibfnamefont {A.}~\bibnamefont {Taoufik}},
  \bibinfo {author} {\bibfnamefont {M.}~\bibnamefont {Bghour}}, \bibinfo
  {author} {\bibfnamefont {H.~E.}\ \bibnamefont {Ouaddi}}, \bibinfo {author}
  {\bibfnamefont {A.}~\bibnamefont {Tirbiyine}}, \bibinfo {author}
  {\bibfnamefont {B.}~\bibnamefont {Lmouden}}, \bibinfo {author} {\bibfnamefont
  {A.}~\bibnamefont {Hafid}},\ and\ \bibinfo {author} {\bibfnamefont {H.~E.}\
  \bibnamefont {Hamidi}},\ }\href {https://doi.org/10.1002/pssb.202100292}
  {\bibfield  {journal} {\bibinfo  {journal} {physica status solidi (b)}\
  }\textbf {\bibinfo {volume} {258}},\ \bibinfo {pages} {2100292} (\bibinfo
  {year} {2021})}\BibitemShut {NoStop}%
\bibitem [{\citenamefont {Prohammer}\ and\ \citenamefont
  {Carbotte}(1991)}]{prohammer1991}%
  \BibitemOpen
  \bibfield  {author} {\bibinfo {author} {\bibfnamefont {M.}~\bibnamefont
  {Prohammer}}\ and\ \bibinfo {author} {\bibfnamefont {J.~P.}\ \bibnamefont
  {Carbotte}},\ }\href {https://doi.org/10.1103/PhysRevB.43.5370} {\bibfield
  {journal} {\bibinfo  {journal} {Phys. Rev. B}\ }\textbf {\bibinfo {volume}
  {43}},\ \bibinfo {pages} {5370} (\bibinfo {year} {1991})}\BibitemShut
  {NoStop}%
\bibitem [{\citenamefont {Basov}\ and\ \citenamefont
  {Timusk}(2005)}]{basov2005}%
  \BibitemOpen
  \bibfield  {author} {\bibinfo {author} {\bibfnamefont {D.~N.}\ \bibnamefont
  {Basov}}\ and\ \bibinfo {author} {\bibfnamefont {T.}~\bibnamefont {Timusk}},\
  }\href {https://doi.org/10.1103/RevModPhys.77.721} {\bibfield  {journal}
  {\bibinfo  {journal} {Rev. Mod. Phys.}\ }\textbf {\bibinfo {volume} {77}},\
  \bibinfo {pages} {721} (\bibinfo {year} {2005})}\BibitemShut {NoStop}%
\bibitem [{\citenamefont {Stilp}\ \emph {et~al.}(2014)\citenamefont {Stilp},
  \citenamefont {Suter}, \citenamefont {Prokscha}, \citenamefont {Salman},
  \citenamefont {Morenzoni}, \citenamefont {Keller}, \citenamefont {Katzer},
  \citenamefont {Schmidl},\ and\ \citenamefont {D\"obeli}}]{stilp2014}%
  \BibitemOpen
  \bibfield  {author} {\bibinfo {author} {\bibfnamefont {E.}~\bibnamefont
  {Stilp}}, \bibinfo {author} {\bibfnamefont {A.}~\bibnamefont {Suter}},
  \bibinfo {author} {\bibfnamefont {T.}~\bibnamefont {Prokscha}}, \bibinfo
  {author} {\bibfnamefont {Z.}~\bibnamefont {Salman}}, \bibinfo {author}
  {\bibfnamefont {E.}~\bibnamefont {Morenzoni}}, \bibinfo {author}
  {\bibfnamefont {H.}~\bibnamefont {Keller}}, \bibinfo {author} {\bibfnamefont
  {C.}~\bibnamefont {Katzer}}, \bibinfo {author} {\bibfnamefont
  {F.}~\bibnamefont {Schmidl}},\ and\ \bibinfo {author} {\bibfnamefont
  {M.}~\bibnamefont {D\"obeli}},\ }\href
  {https://doi.org/10.1103/PhysRevB.89.020510} {\bibfield  {journal} {\bibinfo
  {journal} {Phys. Rev. B}\ }\textbf {\bibinfo {volume} {89}},\ \bibinfo
  {pages} {020510} (\bibinfo {year} {2014})}\BibitemShut {NoStop}%
\bibitem [{\citenamefont {Auslaender}\ \emph {et~al.}(2009)\citenamefont
  {Auslaender}, \citenamefont {Luan}, \citenamefont {Straver}, \citenamefont
  {Hoffman}, \citenamefont {Koshnick}, \citenamefont {Zeldov}, \citenamefont
  {Bonn}, \citenamefont {Liang}, \citenamefont {Hardy},\ and\ \citenamefont
  {Moler}}]{auslaender2009}%
  \BibitemOpen
  \bibfield  {author} {\bibinfo {author} {\bibfnamefont {O.~M.}\ \bibnamefont
  {Auslaender}}, \bibinfo {author} {\bibfnamefont {L.}~\bibnamefont {Luan}},
  \bibinfo {author} {\bibfnamefont {E.~W.~J.}\ \bibnamefont {Straver}},
  \bibinfo {author} {\bibfnamefont {J.~E.}\ \bibnamefont {Hoffman}}, \bibinfo
  {author} {\bibfnamefont {N.~C.}\ \bibnamefont {Koshnick}}, \bibinfo {author}
  {\bibfnamefont {E.}~\bibnamefont {Zeldov}}, \bibinfo {author} {\bibfnamefont
  {D.~A.}\ \bibnamefont {Bonn}}, \bibinfo {author} {\bibfnamefont
  {R.}~\bibnamefont {Liang}}, \bibinfo {author} {\bibfnamefont {W.~N.}\
  \bibnamefont {Hardy}},\ and\ \bibinfo {author} {\bibfnamefont {K.~A.}\
  \bibnamefont {Moler}},\ }\href {https://doi.org/10.1038/nphys1127} {\bibfield
   {journal} {\bibinfo  {journal} {Nat. Phys.}\ }\textbf {\bibinfo {volume}
  {5}},\ \bibinfo {pages} {35} (\bibinfo {year} {2009})}\BibitemShut {NoStop}%
\bibitem [{\citenamefont {Acosta}\ \emph {et~al.}(2019)\citenamefont {Acosta},
  \citenamefont {Bouchard}, \citenamefont {Budker}, \citenamefont {Folman},
  \citenamefont {Lenz}, \citenamefont {Maletinsky}, \citenamefont {Rohner},
  \citenamefont {Schlussel},\ and\ \citenamefont {Thiel}}]{acosta2019color}%
  \BibitemOpen
  \bibfield  {author} {\bibinfo {author} {\bibfnamefont {V.~M.}\ \bibnamefont
  {Acosta}}, \bibinfo {author} {\bibfnamefont {L.~S.}\ \bibnamefont
  {Bouchard}}, \bibinfo {author} {\bibfnamefont {D.}~\bibnamefont {Budker}},
  \bibinfo {author} {\bibfnamefont {R.}~\bibnamefont {Folman}}, \bibinfo
  {author} {\bibfnamefont {T.}~\bibnamefont {Lenz}}, \bibinfo {author}
  {\bibfnamefont {P.}~\bibnamefont {Maletinsky}}, \bibinfo {author}
  {\bibfnamefont {D.}~\bibnamefont {Rohner}}, \bibinfo {author} {\bibfnamefont
  {Y.}~\bibnamefont {Schlussel}},\ and\ \bibinfo {author} {\bibfnamefont
  {L.}~\bibnamefont {Thiel}},\ }\href
  {https://doi.org/10.1007/s10948-018-4877-3} {\bibfield  {journal} {\bibinfo
  {journal} {J. Supercond. Nov. Magn.}\ }\textbf {\bibinfo {volume} {32}},\
  \bibinfo {pages} {85} (\bibinfo {year} {2019})}\BibitemShut {NoStop}%
\bibitem [{\citenamefont {Kazi}\ \emph {et~al.}(2021)\citenamefont {Kazi},
  \citenamefont {Shelby}, \citenamefont {Watanabe}, \citenamefont {Itoh},
  \citenamefont {Shutthanandan}, \citenamefont {Wiggins},\ and\ \citenamefont
  {Fu}}]{kazi2021wide}%
  \BibitemOpen
  \bibfield  {author} {\bibinfo {author} {\bibfnamefont {Z.}~\bibnamefont
  {Kazi}}, \bibinfo {author} {\bibfnamefont {I.~M.}\ \bibnamefont {Shelby}},
  \bibinfo {author} {\bibfnamefont {H.}~\bibnamefont {Watanabe}}, \bibinfo
  {author} {\bibfnamefont {K.~M.}\ \bibnamefont {Itoh}}, \bibinfo {author}
  {\bibfnamefont {V.}~\bibnamefont {Shutthanandan}}, \bibinfo {author}
  {\bibfnamefont {P.~A.}\ \bibnamefont {Wiggins}},\ and\ \bibinfo {author}
  {\bibfnamefont {K.-M.~C.}\ \bibnamefont {Fu}},\ }\href
  {https://link.aps.org/doi/10.1103/PhysRevApplied.15.054032} {\bibfield
  {journal} {\bibinfo  {journal} {Phys. Rev. Appl.}\ }\textbf {\bibinfo
  {volume} {15}},\ \bibinfo {pages} {054032} (\bibinfo {year}
  {2021})}\BibitemShut {NoStop}%
\bibitem [{\citenamefont {Degen}\ \emph {et~al.}(2017)\citenamefont {Degen},
  \citenamefont {Reinhard},\ and\ \citenamefont
  {Cappellaro}}]{degen2017quantum}%
  \BibitemOpen
  \bibfield  {author} {\bibinfo {author} {\bibfnamefont {C.~L.}\ \bibnamefont
  {Degen}}, \bibinfo {author} {\bibfnamefont {F.}~\bibnamefont {Reinhard}},\
  and\ \bibinfo {author} {\bibfnamefont {P.}~\bibnamefont {Cappellaro}},\
  }\href {https://doi.org/10.1103/RevModPhys.89.035002} {\bibfield  {journal}
  {\bibinfo  {journal} {Rev. Mod. Phys.}\ }\textbf {\bibinfo {volume} {89}},\
  \bibinfo {pages} {035002} (\bibinfo {year} {2017})}\BibitemShut {NoStop}%
\end{thebibliography}

\begin{thebibliography}{14}%
\makeatletter
\providecommand \@ifxundefined [1]{%
 \@ifx{#1\undefined}
}%
\providecommand \@ifnum [1]{%
 \ifnum #1\expandafter \@firstoftwo
 \else \expandafter \@secondoftwo
 \fi
}%
\providecommand \@ifx [1]{%
 \ifx #1\expandafter \@firstoftwo
 \else \expandafter \@secondoftwo
 \fi
}%
\providecommand \natexlab [1]{#1}%
\providecommand \enquote  [1]{``#1''}%
\providecommand \bibnamefont  [1]{#1}%
\providecommand \bibfnamefont [1]{#1}%
\providecommand \citenamefont [1]{#1}%
\providecommand \href@noop [0]{\@secondoftwo}%
\providecommand \href [0]{\begingroup \@sanitize@url \@href}%
\providecommand \@href[1]{\@@startlink{#1}\@@href}%
\providecommand \@@href[1]{\endgroup#1\@@endlink}%
\providecommand \@sanitize@url [0]{\catcode `\\12\catcode `\$12\catcode
  `\&12\catcode `\#12\catcode `\^12\catcode `\_12\catcode `\%12\relax}%
\providecommand \@@startlink[1]{}%
\providecommand \@@endlink[0]{}%
\providecommand \url  [0]{\begingroup\@sanitize@url \@url }%
\providecommand \@url [1]{\endgroup\@href {#1}{\urlprefix }}%
\providecommand \urlprefix  [0]{URL }%
\providecommand \Eprint [0]{\href }%
\providecommand \doibase [0]{https://doi.org/}%
\providecommand \selectlanguage [0]{\@gobble}%
\providecommand \bibinfo  [0]{\@secondoftwo}%
\providecommand \bibfield  [0]{\@secondoftwo}%
\providecommand \translation [1]{[#1]}%
\providecommand \BibitemOpen [0]{}%
\providecommand \bibitemStop [0]{}%
\providecommand \bibitemNoStop [0]{.\EOS\space}%
\providecommand \EOS [0]{\spacefactor3000\relax}%
\providecommand \BibitemShut  [1]{\csname bibitem#1\endcsname}%
\let\auto@bib@innerbib\@empty
\bibitem [{\citenamefont {Fujiwara}\ \emph {et~al.}(2020)\citenamefont
  {Fujiwara}, \citenamefont {Dohms}, \citenamefont {Suto}, \citenamefont
  {Nishimura}, \citenamefont {Oshimi}, \citenamefont {Teki}, \citenamefont
  {Cai}, \citenamefont {Benson},\ and\ \citenamefont {Shikano}}]{si_Fujiwara2020}%
  \BibitemOpen
  \bibfield  {author} {\bibinfo {author} {\bibfnamefont {M.}~\bibnamefont
  {Fujiwara}}, \bibinfo {author} {\bibfnamefont {A.}~\bibnamefont {Dohms}},
  \bibinfo {author} {\bibfnamefont {K.}~\bibnamefont {Suto}}, \bibinfo {author}
  {\bibfnamefont {Y.}~\bibnamefont {Nishimura}}, \bibinfo {author}
  {\bibfnamefont {K.}~\bibnamefont {Oshimi}}, \bibinfo {author} {\bibfnamefont
  {Y.}~\bibnamefont {Teki}}, \bibinfo {author} {\bibfnamefont {K.}~\bibnamefont
  {Cai}}, \bibinfo {author} {\bibfnamefont {O.}~\bibnamefont {Benson}},\ and\
  \bibinfo {author} {\bibfnamefont {Y.}~\bibnamefont {Shikano}},\ }\href
  {https://doi.org/10.1103/physrevresearch.2.043415} {\bibfield  {journal}
  {\bibinfo  {journal} {Phys. Rev. Research}\ }\textbf {\bibinfo {volume}
  {2}},\ \bibinfo {pages} {043415} (\bibinfo {year} {2020})}\BibitemShut
  {NoStop}%
\bibitem [{\citenamefont {Ito}\ \emph {et~al.}(2023)\citenamefont {Ito},
  \citenamefont {Tsukamoto}, \citenamefont {Ogawa}, \citenamefont {Teraji},
  \citenamefont {Sasaki},\ and\ \citenamefont {Kobayashi}}]{si_itoh2023}%
  \BibitemOpen
  \bibfield  {author} {\bibinfo {author} {\bibfnamefont {S.}~\bibnamefont
  {Ito}}, \bibinfo {author} {\bibfnamefont {M.}~\bibnamefont {Tsukamoto}},
  \bibinfo {author} {\bibfnamefont {K.}~\bibnamefont {Ogawa}}, \bibinfo
  {author} {\bibfnamefont {T.}~\bibnamefont {Teraji}}, \bibinfo {author}
  {\bibfnamefont {K.}~\bibnamefont {Sasaki}},\ and\ \bibinfo {author}
  {\bibfnamefont {K.}~\bibnamefont {Kobayashi}},\ }\href
  {https://doi.org/10.7566/JPSJ.92.084701} {\bibfield  {journal} {\bibinfo
  {journal} {Journal of the Physical Society of Japan}\ }\textbf {\bibinfo
  {volume} {92}},\ \bibinfo {pages} {084701} (\bibinfo {year}
  {2023})}\BibitemShut {NoStop}%
\bibitem [{\citenamefont {Bezanson}\ \emph {et~al.}(2017)\citenamefont
  {Bezanson}, \citenamefont {Edelman}, \citenamefont {Karpinski},\ and\
  \citenamefont {Shah}}]{julia}%
  \BibitemOpen
  \bibfield  {author} {\bibinfo {author} {\bibfnamefont {J.}~\bibnamefont
  {Bezanson}}, \bibinfo {author} {\bibfnamefont {A.}~\bibnamefont {Edelman}},
  \bibinfo {author} {\bibfnamefont {S.}~\bibnamefont {Karpinski}},\ and\
  \bibinfo {author} {\bibfnamefont {V.~B.}\ \bibnamefont {Shah}},\ }\href
  {https://doi.org/10.1137/141000671} {\bibfield  {journal} {\bibinfo
  {journal} {SIAM Review}\ }\textbf {\bibinfo {volume} {59}},\ \bibinfo {pages}
  {65} (\bibinfo {year} {2017})}\BibitemShut {NoStop}%
\bibitem [{\citenamefont {Haddad}\ and\ \citenamefont {Akansu}(1991)}]{haddad}%
  \BibitemOpen
  \bibfield  {author} {\bibinfo {author} {\bibfnamefont {R.}~\bibnamefont
  {Haddad}}\ and\ \bibinfo {author} {\bibfnamefont {A.}~\bibnamefont
  {Akansu}},\ }\href {https://doi.org/10.1109/78.80892} {\bibfield  {journal}
  {\bibinfo  {journal} {IEEE Transactions on Signal Processing}\ }\textbf
  {\bibinfo {volume} {39}},\ \bibinfo {pages} {723} (\bibinfo {year}
  {1991})}\BibitemShut {NoStop}%
\bibitem [{\citenamefont {Born}\ \emph {et~al.}(1999)\citenamefont {Born},
  \citenamefont {Wolf}, \citenamefont {Bhatia}, \citenamefont {Clemmow},
  \citenamefont {Gabor}, \citenamefont {Stokes}, \citenamefont {Taylor},
  \citenamefont {Wayman},\ and\ \citenamefont {Wilcock}}]{born_wolf}%
  \BibitemOpen
  \bibfield  {author} {\bibinfo {author} {\bibfnamefont {M.}~\bibnamefont
  {Born}}, \bibinfo {author} {\bibfnamefont {E.}~\bibnamefont {Wolf}}, \bibinfo
  {author} {\bibfnamefont {A.~B.}\ \bibnamefont {Bhatia}}, \bibinfo {author}
  {\bibfnamefont {P.~C.}\ \bibnamefont {Clemmow}}, \bibinfo {author}
  {\bibfnamefont {D.}~\bibnamefont {Gabor}}, \bibinfo {author} {\bibfnamefont
  {A.~R.}\ \bibnamefont {Stokes}}, \bibinfo {author} {\bibfnamefont {A.~M.}\
  \bibnamefont {Taylor}}, \bibinfo {author} {\bibfnamefont {P.~A.}\
  \bibnamefont {Wayman}},\ and\ \bibinfo {author} {\bibfnamefont {W.~L.}\
  \bibnamefont {Wilcock}},\ }\href {https://doi.org/10.1017/CBO9781139644181}
  {\emph {\bibinfo {title} {Principles of Optics: Electromagnetic Theory of
  Propagation, Interference and Diffraction of Light}}},\ \bibinfo {edition}
  {7th}\ ed.\ (\bibinfo  {publisher} {Cambridge University Press},\ \bibinfo
  {year} {1999})\BibitemShut {NoStop}%
\bibitem [{\citenamefont {Nishimura}\ \emph {et~al.}()\citenamefont
  {Nishimura}, \citenamefont {Tsukamoto}, \citenamefont {Sasaki},\ and\
  \citenamefont {Kobayashi}}]{si_nishimura2023tobesubmitted}%
  \BibitemOpen
  \bibfield  {author} {\bibinfo {author} {\bibfnamefont {S.}~\bibnamefont
  {Nishimura}}, \bibinfo {author} {\bibfnamefont {M.}~\bibnamefont
  {Tsukamoto}}, \bibinfo {author} {\bibfnamefont {K.}~\bibnamefont {Sasaki}},\
  and\ \bibinfo {author} {\bibfnamefont {K.}~\bibnamefont {Kobayashi}},\
  }\href@noop {} {}\bibinfo {note} {In preparation}\BibitemShut {NoStop}%
\bibitem [{\citenamefont {Carneiro}\ and\ \citenamefont
  {Brandt}(2000)}]{si_carneiro2000vortex}%
  \BibitemOpen
  \bibfield  {author} {\bibinfo {author} {\bibfnamefont {G.}~\bibnamefont
  {Carneiro}}\ and\ \bibinfo {author} {\bibfnamefont {E.~H.}\ \bibnamefont
  {Brandt}},\ }\href {https://doi.org/10.1103/PhysRevB.61.6370} {\bibfield
  {journal} {\bibinfo  {journal} {Phys. Rev. B}\ }\textbf {\bibinfo {volume}
  {61}},\ \bibinfo {pages} {6370} (\bibinfo {year} {2000})}\BibitemShut
  {NoStop}%
\bibitem [{\citenamefont {Kogan}(2003)}]{si_kogan2003}%
  \BibitemOpen
  \bibfield  {author} {\bibinfo {author} {\bibfnamefont {V.~G.}\ \bibnamefont
  {Kogan}},\ }\href {https://doi.org/10.1103/PhysRevB.68.104511} {\bibfield
  {journal} {\bibinfo  {journal} {Phys. Rev. B}\ }\textbf {\bibinfo {volume}
  {68}},\ \bibinfo {pages} {104511} (\bibinfo {year} {2003})}\BibitemShut
  {NoStop}%
\bibitem [{\citenamefont {Kogan}\ \emph {et~al.}(2021)\citenamefont {Kogan},
  \citenamefont {Nakagawa},\ and\ \citenamefont {Kirtley}}]{si_kogan2021}%
  \BibitemOpen
  \bibfield  {author} {\bibinfo {author} {\bibfnamefont {V.~G.}\ \bibnamefont
  {Kogan}}, \bibinfo {author} {\bibfnamefont {N.}~\bibnamefont {Nakagawa}},\
  and\ \bibinfo {author} {\bibfnamefont {J.~R.}\ \bibnamefont {Kirtley}},\
  }\href {https://doi.org/10.1103/PhysRevB.104.144512} {\bibfield  {journal}
  {\bibinfo  {journal} {Phys. Rev. B}\ }\textbf {\bibinfo {volume} {104}},\
  \bibinfo {pages} {144512} (\bibinfo {year} {2021})}\BibitemShut {NoStop}%
\bibitem [{\citenamefont {Davis}\ and\ \citenamefont
  {Rabinowitz}(2007)}]{davis2007}%
  \BibitemOpen
  \bibfield  {author} {\bibinfo {author} {\bibfnamefont {P.~J.}\ \bibnamefont
  {Davis}}\ and\ \bibinfo {author} {\bibfnamefont {P.}~\bibnamefont
  {Rabinowitz}},\ }\href {https://doi.org/10.1016/C2013-0-10566-1} {\emph
  {\bibinfo {title} {Methods of numerical integration}}}\ (\bibinfo
  {publisher} {Courier Corporation},\ \bibinfo {year} {2007})\BibitemShut
  {NoStop}%
\bibitem [{\citenamefont {Brent}(1971)}]{brent}%
  \BibitemOpen
  \bibfield  {author} {\bibinfo {author} {\bibfnamefont {R.}~\bibnamefont
  {Brent}},\ }\emph {\bibinfo {title} {Algorithms for finding zeros and extrema
  of functions without calculating derivatives}},\ \href@noop {} {Ph.D.
  thesis},\ \bibinfo  {school} {Stanford University} (\bibinfo {year}
  {1971})\BibitemShut {NoStop}%
\bibitem [{\citenamefont {Transtrum}\ \emph {et~al.}(2011)\citenamefont
  {Transtrum}, \citenamefont {Machta},\ and\ \citenamefont
  {Sethna}}]{transtrum2011}%
  \BibitemOpen
  \bibfield  {author} {\bibinfo {author} {\bibfnamefont {M.~K.}\ \bibnamefont
  {Transtrum}}, \bibinfo {author} {\bibfnamefont {B.~B.}\ \bibnamefont
  {Machta}},\ and\ \bibinfo {author} {\bibfnamefont {J.~P.}\ \bibnamefont
  {Sethna}},\ }\href {https://doi.org/10.1103/PhysRevE.83.036701} {\bibfield
  {journal} {\bibinfo  {journal} {Phys. Rev. E}\ }\textbf {\bibinfo {volume}
  {83}},\ \bibinfo {pages} {036701} (\bibinfo {year} {2011})}\BibitemShut
  {NoStop}%
\bibitem [{\citenamefont {Hansen}\ \emph {et~al.}(2013)\citenamefont {Hansen},
  \citenamefont {Pereyra},\ and\ \citenamefont {Scherer}}]{hansen_lsq}%
  \BibitemOpen
  \bibfield  {author} {\bibinfo {author} {\bibfnamefont {P.~C.}\ \bibnamefont
  {Hansen}}, \bibinfo {author} {\bibfnamefont {V.}~\bibnamefont {Pereyra}},\
  and\ \bibinfo {author} {\bibfnamefont {G.}~\bibnamefont {Scherer}},\
  }\href@noop {} {\emph {\bibinfo {title} {Least Squares Data Fitting with
  Applications}}}\ (\bibinfo  {publisher} {Johns Hopkins University Press},\
  \bibinfo {year} {2013})\BibitemShut {NoStop}%
\bibitem [{\citenamefont {Tsukamoto}\ \emph {et~al.}(2021)\citenamefont
  {Tsukamoto}, \citenamefont {Ogawa}, \citenamefont {Ozawa}, \citenamefont
  {Iwasaki}, \citenamefont {Hatano}, \citenamefont {Sasaki},\ and\
  \citenamefont {Kobayashi}}]{si_tsukamoto2021vector}%
  \BibitemOpen
  \bibfield  {author} {\bibinfo {author} {\bibfnamefont {M.}~\bibnamefont
  {Tsukamoto}}, \bibinfo {author} {\bibfnamefont {K.}~\bibnamefont {Ogawa}},
  \bibinfo {author} {\bibfnamefont {H.}~\bibnamefont {Ozawa}}, \bibinfo
  {author} {\bibfnamefont {T.}~\bibnamefont {Iwasaki}}, \bibinfo {author}
  {\bibfnamefont {M.}~\bibnamefont {Hatano}}, \bibinfo {author} {\bibfnamefont
  {K.}~\bibnamefont {Sasaki}},\ and\ \bibinfo {author} {\bibfnamefont
  {K.}~\bibnamefont {Kobayashi}},\ }\href {https://doi.org/10.1063/5.0054809}
  {\bibfield  {journal} {\bibinfo  {journal} {Appl. Phys. Lett.}\ }\textbf
  {\bibinfo {volume} {118}},\ \bibinfo {pages} {264002} (\bibinfo {year}
  {2021})}\BibitemShut {NoStop}%
\end{thebibliography}
\end{document}